\documentclass[reprint,aps,prx,amsmath,amssymb,longbibliograaphy, nofootinbib, notitlepage]{revtex4-1}
\usepackage{amsmath,amssymb,bm,graphicx}
\usepackage{graphics}
\usepackage{float} 
\usepackage[stable]{footmisc}
\usepackage{soul}
\usepackage[colorlinks, linkcolor= blue, citecolor = blue, urlcolor=blue]{hyperref}

\def\be{\begin{equation}}
\def\ee{\end{equation}}
\def \bea{\begin{eqnarray}}
\def \eea{\end{eqnarray}}
\def \nn{\nonumber}

\begin{document}

%\title{Nonlinear magnetotransport in Weyl semimetal in quantizing magnetic field}
\title{Nonlinear magnetoconductivity in Weyl and multi-Weyl semimetal in quantizing magnetic field}
\author{Sunit Das}
\email{sunitd@iitk.ac.in}
\author{Kamal Das}
\email{kamaldas@iitk.ac.in}
\author{Amit Agarwal}
\email{amitag@iitk.ac.in}
\affiliation{Department of Physics, Indian Institute of Technology, Kanpur-208016, India}
\thanks{S.D. and K.D. contributed equally to this work.}

\begin{abstract}
Magnetotransport and magneto-optics experiments offer a very powerful probe for studying the physical properties of materials. Here, we investigate the second-order nonlinear magnetoconductivity of tilted type-I Weyl and multi-Weyl semimetal. 
In contrast to the presence of chiral anomaly in the linear response regime, we show that Weyl semimetal do not host chiral charge pumping in the nonlinear transport regime.
We predict that an inversion symmetry broken and tilted Weyl semimetal can support finite longitudinal nonlinear magnetoconductivity, which is otherwise absent in untilted Weyl semimetal. The nonlinear magnetoconductivity vanishes in the ultra-quantum limit, oscillates in the intermediate magnetic field regime and saturates in the semiclassical limit. The nonlinear magnetoconductivity depends intricately on the tilt orientation, and it can be used to determine the tilt orientation in Weyl and multi-Weyl semimetals, via nonlinear magnetoresistivity or second harmonic generation experiments.
 
\end{abstract}
\maketitle

%\section{Introduction}
%%%%%{\bf general background of WSMs} %%%%%%%%

{\it Introduction:---} 
Since their discovery, Weyl semimetals (WSM) have attracted significant attention due to their unusual linear quasiparticle dispersion mimicking the Weyl fermions with novel topological properties~\cite{wan_PRB2011_topo,burkov_PRL2011_weyl, hasan_ARCM2017_disco,yan_ARCM2017_topo,armitage_RMP2018_weyl,lv_RMP2021_expt}. The combination of Weyl points in the bulk, which act as source and sink of Berry curvature with topological charges, and nontrivial Fermi arc surface states, support diverse novel transport and optical phenomena including the quantum anomalies~\cite{nielsen_PLB1981_the,zyuzin_PRB2012_topo,son_PRB2013_chiral,kim_PRL2013_dirac,deng_PRL2019_posi,das_PRR2020_thermal,das_PRR2020_chiral,chernodub_arxiv2021_thermal,das_PRB2021_intrinsic}. Several of these phenomena has been experimentally realized in a wide range of materials showing Weyl characteristics, starting from the three dimensional Dirac semimetal in presence of a magnetic field~\cite{li_NC2016_nega,xiong_S2015_eviden} to transition metal mononictides~\cite{weng_PRX_weyl,lv_PRX2015_expt,xu_S2015_disco,huang_NC2015_a} and magnetic materials~\cite{zyuzin_PRB2012_weyl,chang_PRB2018_magne,yang_PRB2021_non}. Furthermore, the realization of WSM in space inversion symmetry (SIS) broken systems has facilitated the   exploration and potential application of second order nonlinear (NL) responses~\cite{Boyd20,Wu17, ma_NP2017_dir}. It has been shown that the SIS broken WSM exhibit photogalvanic responses~\cite{chan_PRB2017_photo, konig_PRB2017_photo, golub_JETPL2017_photo, sadhukhan_prb21_role, sadhukhan_prb21_electronic} like injection current~\cite{dejuan_NC2017_quan} and shift current~\cite{yang_arxiv2018_div}, second harmonic generation~\cite{li_PRB2018_sec,Gao_opt21}, sum and difference frequency generation~\cite{dejuan_PRR2020_diff}, and the nonlinear Hall effect~\cite{sodemann_PRL2015_quan, rostami_PRB2018_non, gao_PRB2020_sec} among others. 

%%%%%{\bf Non-linear magneto-opti%cs in WSM}

Recently, there have been several studies focusing on how the NL transport and optical responses in WSM are modified in the presence of a magnetic field. Treating the magnetic field within the semiclassical framework, it has been shown that the Berry curvature induces a finite NL magnetoconductivity~\cite{morimoto_PRB2016_semi, zyuzin_PRB2018_nonlinear, zyuzin_PRB2018_chiral,li_PRB2021_non, zeng_prb2022_chiral, tanay_arxiv22_distinct}. In  addition, the inter-band transition of carriers, combined with the chiral magnetic velocity gives rise to the helical magnetic effect~\cite{kharzeev_PRB2018_giant}. On the other hand, in presence of a strong magnetic field where quantized Landau levels (LLs) are formed, optical transitions have been shown to generate photocurrent~\cite{golub_JETPL2017_photo, golub_PRB2018_circ} in gyrotropic WSM. More recently, second harmonic generation has been demonstrated in isotropic WSM, in presence of a DC electric field \cite{gao_PRB2021_cur}. %which serves as a tool to break the SIS.

\begin{figure}
    \centering
    \includegraphics[width=\linewidth]{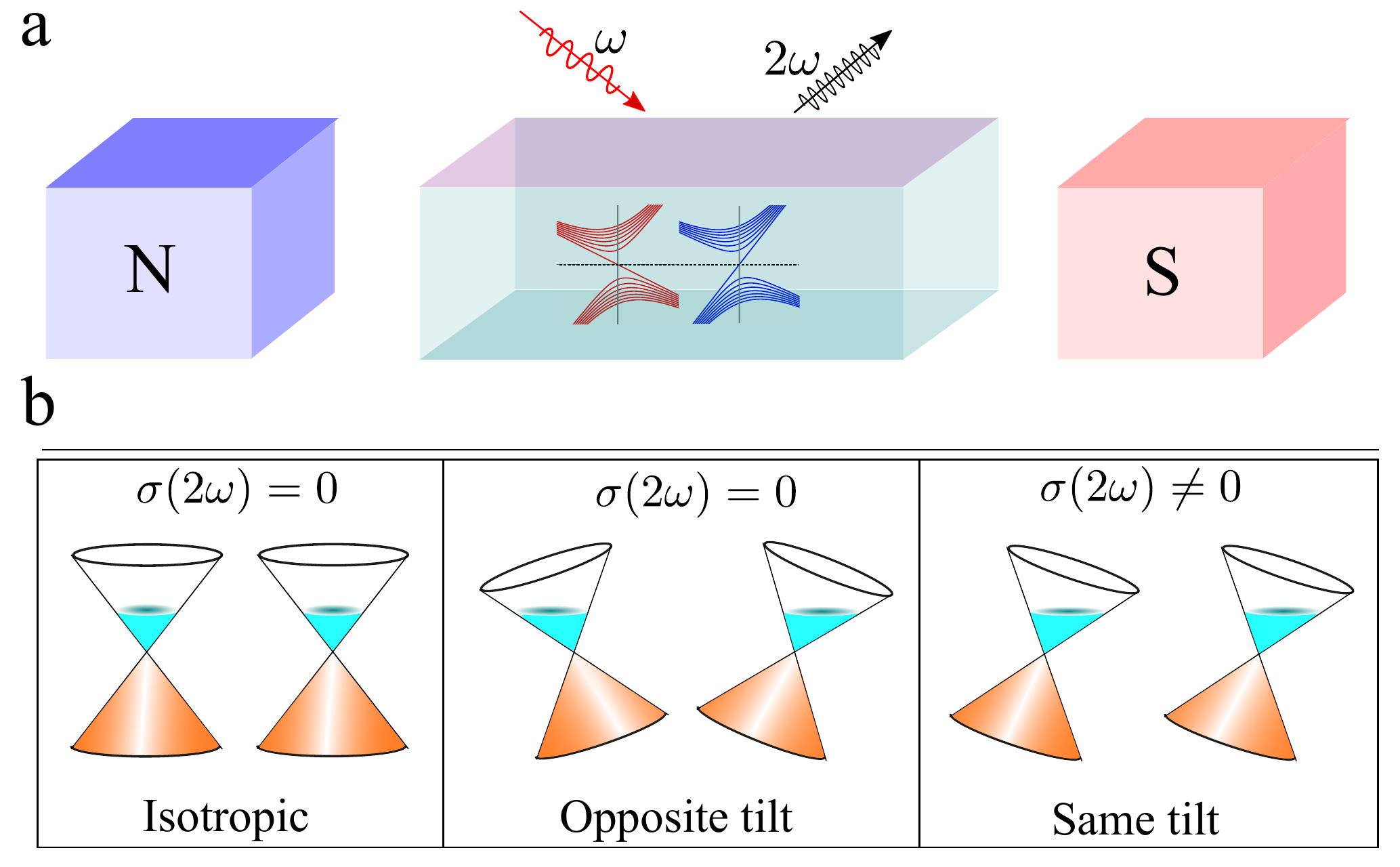}
    \caption{a) A schematic of the second harmonic generation [$\sigma({2 \omega})$] in presence of a quantizing magnetic field. Panel b) presents a summary of the various tilt orientations in WSM, and the corresponding nonlinear responses. We show that the nonlinear longitudinal response is finite only when both the space inversion symmetry and time reversal symmetry in WSM are broken, with the tilt direction in the Weyl nodes of opposite chirality being aligned with each other.}
    \label{fig_1}
\end{figure}

%%%%{\bf summary of the paper and its novelty}

Motivated by these exciting studies, in this paper we explore the second-order NL magnetotransport/optical response in WSM and in multi-WSM~\cite{li_PRB2016_weyl, gupta_PLA2019_novel, dantas_JHEP2018_mag, menon_PRB2021_chiral, mukherjee_PRB2018_dop, nandy_PRB2021_chiral, nag_JPCM2020_magneto, fu_PRB2022_thermo, xiong_arxiv22_understanding, anirudha_arxiv19_anomalous,tanay_prb20_thermoelectric} in presence of a strong magnetic field that gives rise to discrete LLs [see Fig.~\ref{fig_1}(a)]. We ask the fundamental question: do the WSM support NL chiral charge pumping that is second order in electric field, similar to the linear chiral anomaly~\cite{nielsen_PLB1981_the,son_PRB2013_chiral}, %the chiral charge pumping from one Weyl node to the other 
which is proportional to the electric field strength. %, and if yes then how do that manifest through the NL conductivities. 
To address this, and to treat the semiclassical, quantum oscillation, and the ultra-quantum transport regimes on the same footing, we apply the Boltzmann transport framework to the quantized LLs.
Specifically, we calculate the intra-band contribution to the NL longitudinal conductivity 
$\sigma_{zzz}$, which relates the second-order NL current to the applied electric fields, $j_z=\sigma_{zzz}E_z E_z$ [see Eq.~\eqref{sigma}]. 

We consider a low energy model Hamiltonian describing a pair of tilted Weyl nodes of opposite chirality ($\chi=\pm 1$), specified by~\cite{sonowal_PRB2019_giant,das_PRB2019_linear,das_PRB2019_berry}
\be  \label{ham}
\mathcal{H}_\chi= \chi  v_F \hbar {\bm k} \cdot {\bm \sigma}+ \hbar{\bm w}_\chi \cdot {\bm k} ~ \sigma_0.
\ee 
Here, $v_F$ is the Fermi velocity, ${\bm \sigma}=(\sigma_{x},\sigma_y, \sigma_z)$ denotes the Pauli matrices, $\sigma_0$ is the identity matrix and ${\bm w}_\chi $ represents the tilt velocity.  
We find that for having a finite NL longitudinal response, in addition to the broken SIS the time reversal symmetry (TRS) also needs to be broken. Different possible tilt orientations for the pair of opposite chirality Weyl nodes, and the resulting NL longitudinal conductivity is summarized in Fig.~\ref{fig_1}(b). We demonstrate that i) WSM do not host second order NL chiral anomaly {\it i.e.}, there is no chiral charge pumping that is  quadratic in electric field, and ii) consequently the whole NL longitudinal conductivity is determined by the intra-node scattering times. We show that the NL longitudinal conductivity vanishes in the ultra-quantum limit indicating that the chiral LL do not contribute to it. The NL conductivity shows an oscillating behaviour in $1/B$ in the intermediate magnetic field regime, and the period of quantum oscillations can be used to measure the tilt velocity. In the semiclassical limit, the NL conductivity becomes independent of $B$ reducing to the NL counterpart of the Drude conductivity. Our calculations reveal that these features of NL longitudinal response also persist in  multi-WSM.

{\it Symmetries and tilt orientation in Weyl semimetal:---} The TRS and SIS play a fundamental role in determining the low energy Hamiltonian of tilted WSM~\cite{yu_PRL2016_pred,konig_PRB2017_photo, armitage_RMP2018_weyl,chen_PRB2019_optical}, and consequently the NL responses. %In WSM, Weyl nodes always comes in pairs of opposite chirality. 
In space inversion symmetric (TRS broken) WSM, minimum two nodes are feasible and the Weyl nodes of opposite chirality are related via the centre of inversion~\cite{armitage_RMP2018_weyl}.
In such systems, if the low energy model of one Weyl node in the Brillouin zone is given by ${\mathcal H}={\bm \sigma} \cdot {\bm k} + {\bm w}\cdot {\bm k}$, 
%($\bm \sigma$ denotes the vector of Pauli matrices, and $\bm w$ represents the direction of the tilt), 
then the Hamiltonian of the other node related through SIS is obtained by ${\bm k} \to - {\bm k}$ as ${\mathcal H}_{\mathcal P}=-{\bm \sigma} \cdot {\bm k} - {\bm w}\cdot {\bm k}$. This indicates that the SIS related Weyl nodes will always be oppositely tilted (${\bm w}_- = -{\bm w}_+)$.
On the other hand, in a TRS preserving (SIS broken) WSM, a minimum of four nodes are needed. The Weyl nodes of the same chirality are related through a time-reversal invariant momentum while there is no symmetry restriction between the nodes with opposite chirality~\cite{armitage_RMP2018_weyl}. In this case, if the low energy model of one of the Weyl node is given by ${\mathcal H}={\bm \sigma} \cdot {\bm k} + {\bm w}\cdot {\bm k}$, then the Hamiltonian of the other same chirality node related to it via TRS is obtained to be ${\mathcal H}_{\mathcal T}={\bm \sigma} \cdot {\bm k} - {\bm w}\cdot {\bm k}$~\cite{yu_PRL2016_pred}. This implies that nodes with the same chirality have opposite tilt.
Since the second-order NL response vanishes in presence of SIS, we consider the case of SIS broken WSM, in which the Weyl nodes of opposite chirality have the same tilt velocity (${\bm w}_+ = {\bm w}_-$). Within the family of SIS broken WSM, we can consider systems either with TRS or without TRS. We show later that in TRS preserving WSM, the longitudinal NL response vanishes [see Eq.~\eqref{SHG_int_tilted}]. 
%The reason for this will be discussed later. 
Therefore, our work is focused on WSM without any fundamental symmetries (both TRS and SIS are broken).

%\section{Landau Levels in tilted Weyl semimetals}
%\label{LLs}
%%%%{\bf preparing the stage} %%%%%%%

{\it Landau levels in tilted Weyl semimetal:---} The eigenvalue problem of massless tilted Dirac fermions in presence of a strong magnetic field has been earlier explored in three dimensional~\cite{yu_PRL2016_pred, shao_JPCM2021_long} systems as well as in two dimensional~\cite{goerbig_EPL2009_elec, sari_PRB2015_mag} systems. Here, we sketch the calculation for a three dimensional Weyl Hamiltonian given in Eq.~\eqref{ham}. 
%, for the sake of completeness. 
To be specific, we consider the magnetic field to be applied along the $z$-axis (${\bm B} = B {\hat z}$), 
and the tilt velocity to be in the $x$-$z$ plane, ${\bm w}_{\chi} =( w_{\bot,\chi}, 0, w_{\parallel,\chi})$ having components parallel to and perpendicular to the applied magnetic field. We define the dimensionless quantities i) ${\bm t}_{\chi}= {\bm w}_{\chi}/v_F = ( t_{\bot,\chi},0, t_{\parallel,\chi})=|{\bm t}_\chi|({\sin \theta, 0, \cos \theta})$, $\theta$ being the angle between the ${\bm w}_\chi$ and $\bm B$, and ii) $\alpha= (1-t_{\bot,\chi}^{2})^{1/2}$, which will be used later. Depending on the dimensionless tilt strength, WSM are categorized into two classes. For ${|{\bm t}_\chi|}<1$, the Fermi surface of each Weyl node hosts only a closed electron or a  closed hole pocket (in absence of magnetic field) and such systems are called type-I WSM. The systems with ${|{\bm t}_\chi|}> 1$, are called type-II WSM, and in these systems the Fermi surface (in absence of magnetic field) at the charge neutrality point consists of an open electron and an open hole pocket with the Weyl node connecting the two. 
In this paper, we focus on type-I WSM.

%{\bf sketch of the derivation of LLs and some important points }
%
\begin{figure}
    \centering
    \includegraphics[width=\linewidth]{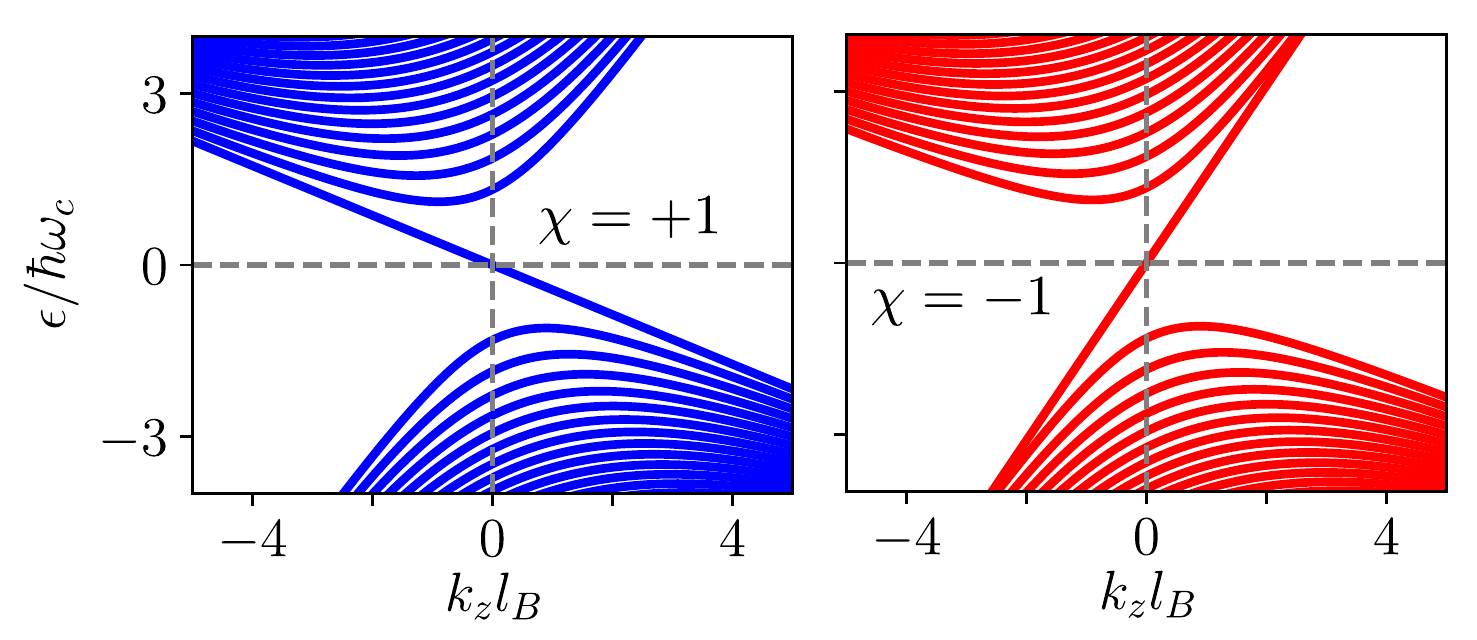}
    \caption{Landau level spectrum of tilted type-I Weyl nodes with chirality $\chi=+1$ (left panel) and $\chi=-1$ (right panel). The energy axis is scaled with $\hbar \omega_c$ and $k_z$ axis is scaled with magnetic length $l_B$. The tilt has been considered to be oriented in same direction for both Weyl nodes of opposite chirality. We have used ${\bm t}_\chi=0.6(\sin \theta, 0 , \cos \theta)$ with $\theta=\pi/6$, Fermi velocity $v_F=2\times 10^{5}$ m/s and magnetic field $B=10$ Tesla.}
    \label{fig_2}
\end{figure}

To calculate the LLs, we use the Landau gauge to represent the magnetic field  in the $z$-direction via the vector potential ${\bm A}=(-By, 0, 0)$. In presence of the vector potential, the Peierls substitution transforms the Hamiltonian as $\mathcal{H}_\chi= \chi  v_F (\hbar \hat {\bm k}+e {\bm A})\cdot {\bm \sigma}+  {\bm w}_\chi \cdot  (\hbar \hat {\bm k} +e {\bm A} ) \sigma_0$. The choice of the Landau gauge breaks the translational symmetry of the system along the $y$-axis, but $k_x$ and $k_z$ remain good quantum numbers. Furthermore, the tilt in the Hamiltonian combined with the vector potential introduces a term like $-e y w_{\bot,\chi} B$ which can be seen as a potential resulting from an effective electric field $E_{\rm eff} = w_{\bot,\chi} B$ along the negative $y$-direction. Importantly, the LLs can exist only when the effective drift velocity, $v_d=E_{\rm eff}/B$ is less than the Fermi velocity i.e.,  $w_{\bot,\chi} < v_F$ ~\cite{Lukose07}. Now, introducing a Lorentz boost to eliminate the $E_{\rm eff}$ field, and after some little algebra we obtain the energy spectrum to be \cite{yu_PRL2016_pred, shao_JPCM2021_long},
\be \label{dispersion}
\epsilon_n^\chi=
\begin{cases}
(-\chi \alpha + t_{\parallel,\chi}) \hbar v_F k_z & n=0,\\[2ex]
{\rm sgn}(n) \alpha {\mathcal E}_{n,k_z} +  \hbar v_F t_{\parallel,\chi} k_z & n\neq0.
\end{cases}
\ee
Here, we have defined $\omega_c=v_F/l_B$ with $l_B=\sqrt{\hbar/(eB)}$ and ${\mathcal E}_{n,k_z} =\sqrt{2|n|\alpha (\hbar \omega_c)^2 + (\hbar v_F  k_z)^2}$. After Landau quantization, the three-dimensional problem effectively becomes a one-dimensional problem along the $k_z$-axis. The LLs for a pair of Weyl nodes with opposite chirality and the same tilt velocity, are shown in Fig.~\ref{fig_2}. Some  interesting facts about the LLs spectrum of tilted Weyl nodes are as follows: i) For the lowest LL, the tilt only modifies the strength of the band velocity while for higher LLs the tilt introduces an additional term in the  dispersion which is an odd function of $k_z$. We show below that the latter significantly modifies the NL conductivity. 
ii) The tilt displaces the minima of the non-chiral LLs in the $k_z$-axis and also squeezes their dispersion. The energy separation between consecutive LLs (at the minima) is given by $\epsilon_{n+1}-\epsilon_n=[\alpha (1-t_{\chi}^2) ]^{1/2}\hbar \omega_c \sqrt{2(|n + 1| - |n|)}$, which is $[\alpha (1-t_{\chi}^2)]^{1/2}$ times smaller than that in isotropic WSM. iii) The tilt removes the particle-hole symmetry which is otherwise present in isotropic Weyl nodes. iv) The tilt does not alter the degeneracy of energy levels and it is the same as in isotropic WSM, $\mathfrak{D}=1/{2\pi l_B^2}$ per unit cross-section area perpendicular to the magnetic field~\cite{shao_JPCM2021_long}. 

It is straightforward to calculate the density of states,  $\rho(\epsilon)$. For $\mu>0$, we obtain $\rho(\epsilon)$ to be, 
\be \label{DOS}
\rho(\epsilon) =\rho_0 \bigg[ \frac{1}{\alpha-\chi  t_{\parallel,\chi}} +
\sum_{j=\pm, n=1}^{n_c}   \Big(\Big|t_{\parallel,\chi}+ \frac{\hbar v_F \alpha k_{z0}^{j} }{{\mathcal E}_{n,k_{z0}^{j}}} \Big|\Big)^{-1} \bigg].
\ee
Here, we have defined $\rho_0=1/(4 \pi^2 l_B^2 \hbar v_F)$, and $k_{z0}^{j}$ are the momentum points where the constant energy line ($\epsilon$) intersects the LLs. The $k_{z0}^{j}~(j=\pm)$ are given by 
\be \label{momentum_cut} 
k_{z0}^{\pm} (\epsilon) =  \frac{\pm \sqrt{  (1-t_\chi^2) \left( k^2 - 2 |n| \alpha^3 /l_B^2 \right)+ k^2 t_{\parallel,\chi}^2} - k t_{\parallel,\chi}} {1-t_\chi^2},
\ee  
with $k=\epsilon/\hbar v_F$. The DOS in Eq.~\eqref{DOS} is for the conduction band side ($\mu>0$) and for more general form, we refer readers to Ref.~\cite{vadnais_PRB2021_quan}. For each of the LLs, the density of states diverges at $\epsilon_n = \sqrt{2|n| \alpha (1-t_\chi^2)}  \hbar \omega_c$. %which has been earlier obtained in  Ref.~\cite{shao_JPCM2021_long}.
The group velocity along the $z$-direction is calculated to be
\be 
v_{z,n}^\chi= \begin{cases}
(-\chi \alpha + t_{\parallel,\chi}) v_F & n=0,\\
{\rm sgn}(n) \alpha  \hbar v_F^2 k_z/ {\mathcal E}_{n,k_z} + v_F t_{\parallel, \chi} & n \neq 0.
\end{cases}
\ee 
The tilt introduces a constant velocity in each LL.  The LL DOS and the group velocities will be used later to calculate the NL longitudinal conductivity.

{\it Vanishing nonlinear chiral anomaly:---}
In the Boltzmann transport formalism, the current is calculated via the equation, $j(t) = -e {\mathfrak D}\sum_{n, \chi}  \int [dk_z] v_{z,n}^\chi f_n^\chi(t)$. Here, `$-e$' denotes the electronic charge, $[dk_z] \equiv  dk_z/(2 \pi)$ and $v_{z,n}^\chi$ is the velocity along ${z}$-direction and $f_n^\chi(t)$ is the non-equilibrium distribution function (NDF) in presence of the applied external fields. To calculate the NL longitudinal conductivity, we consider a spatially uniform electric field oscillating at frequency $\omega$ and applied parallel to the magnetic field, ${\bm E}(t) = \hat{\bm z}{ E_z} e^{- i \omega t}$. 
In the linear response regime, parallel electric and magnetic field configuration (finite ${\bm E}\cdot{\bm B}$) induces chiral anomaly~\cite{nielsen_PLB1981_the}, the non-conservation of chiral charge in WSM. The chiral charge pumping is countered by inter-node scattering to establish a steady state. Incorporating this in the Boltzmann equation, we can calculate the NDF~\cite{son_PRB2013_chiral}. 

Using the  relaxation time approximation
\cite{deng_PRL2019_posi,deng_PRB2019_modu,shao_JPCM2021_long}, we have
\be\label{BTE}
\partial_t f_n^\chi(t) + \dot{\bm k}_n^\chi \cdot {\bm \nabla}_{\bm k} f_n^\chi(t) =-\frac{f_n^\chi(t) -\Bar{f}_n^\chi(t)}{\tau}-\frac{{\Bar f}_n^\chi(t)-f_n^0}{\tau_v}~.
\ee
Here, $\bar f_n^\chi(t)$ represents the `local equilibrium' distribution function for each Weyl node.
The global equilibrium distribution function is defined as $f_n^0 = [\bar f_n^\chi(t) + \bar f_n^{-\chi}(t)]/2$, which we assume to be the Fermi function, $f_n^0 = 1/[ 1+e^{\beta (\epsilon_n^\chi -\mu)}]$ at chemical potential $\mu$ and inverse temperature $\beta=1/(k_B T)$, with $T$ and $k_B$ being the temperature and the Boltzmann constant, respectively. The first term in the right-hand side of Eq.~(\ref{BTE}) represents the collision integral for the intra-node scattering (with scattering rate $1/{\tau}$), which establishes the local equilibrium. The intra-node scattering does not change the number of carriers in the respective node. The collision integral for inter-node scattering is represented by the second term in Eq.~(\ref{BTE}) with the inter-node scattering rate $1/{\tau_v}$. For simplicity, we ignore the energy dependence of both the scattering time. 
The NDF can be expressed as a sum of the equilibrium and non-equilibrium part by expanding it in powers of the electric field strength, $f_n^\chi(t) = f_n^0 + f_n^{(1),\chi} e^{-i\omega t}+f_n^{(2),\chi} e^{-i 2\omega t} + \cdots $. Here,  $f_n^{(1),\chi}$ is linear order in the electric field, $f_n^{(2),\chi}$ is quadratic order in the electric field ($\propto |{\bm E}|^2$) and so on. In this paper, we are interested in second-order response and hence we focus on calculating  $f_n^{(2),\chi}$. The first question we address is the possibility of having NL chiral charge pumping in WSM in  which the rate change of  chiral charge carriers will be  proportional to $|{\bm E}|^2$. 

%Interestingly, we show below that in contrast to the linear response regime, there are no NL counterpart of the chiral anomaly, {\it i.e.} there is no chiral charge pumping which is proportional to second order in the electric field. Consequently $f_n^{(2),\chi}(t)$ is determined only by the intranode scattering time. 
To explore NL chiral anomaly in WSM, we start by reviewing the linear chiral anomaly and build the second-order response on top of that. The existence of linear chiral anomaly can be deduced from the collisionless Boltzmann equation [Eq.~\eqref{BTE} with $\tau_v$ and $\tau \to \infty$], by using the equilibrium distribution function in the $\dot{{\bm k}}\cdot {\bm \nabla }_{\bm k} f_n^\chi$ term and constructing a continuity equation \cite{zyuzin17_chiral, das_PRR2020_chiral}.  
Integrating over all the momentum states, we obtain \cite{das_PRR2020_chiral}    
\bea \label{linear_CA}
\frac{\partial {\cal  N}^{(1),\chi}}{\partial t} &=& -{\mathfrak D}\sum_n e E_z \int [d k_z] v_{n,z}^\chi \left(-\partial_{\epsilon} f_n^0 \right) \nonumber \\
& = & \frac{\chi e^2}{4 \pi^2 \hbar^2}  E_z B~.
\eea
Here, ${\cal N}^{(1),\chi} = \mathfrak{D} \sum_n \int [dk_z] [f_{n}^{(0),\chi}+f_{n}^{(1),\chi}]$ is the particle number density in each Weyl node. Clearly, the chiral charge density is not conserved and this chiral charge pumping $\propto {\bm E}\cdot {\bm B}$ is the hallmark of linear chiral anomaly~\cite{son_PRB2013_chiral, das_PRR2020_chiral}. 
% the first order response the non-equilibrium part has been expressed as a product of derivative of Fermi function and the remaining factors ($\delta g_n^{(11),\chi}$) as $f_n^{(1),\chi}= \left(- \partial_{\epsilon} f_n^0 \right) \delta g_n^{(11),\chi}$. The first superscript represents the order of electric field and the second one represents the order of Fermi function derivative. Using this approach the first order in electric field contribution to the NDF, $\delta g_n^{(11),\chi}$ is given by \cite{deng_PRL2019_posi, deng_PRB2019_modu, shao_JPCM2021_long, das_PRR2020_chiral}
%
%\bea
%\delta g_n^{(11),\chi} =-e \tau_{\omega} E_z v_{z,n}^\chi  + \kappa \dfrac{\tau_{\omega}}{\tau_v} \langle \delta g_n^{(11),\chi} \rangle ,
%\eea 
%
% we have defined $\langle \delta g_n^{(11),\chi} \rangle = -e \tau_{v,\omega} E_z \langle v_{z,n}^\chi \rangle$. Also, we have denoted
%$\tau_\omega=1/(1- i \omega \tau)$, $\tau_{v,\omega}=1/(1- i \omega \tau_v)$ and $\kappa=({\tau_v/\tau} -1)$. the local equilibrium part of the distribution function is calculated keeping in mind the particle conservation in the from the Boltzmann equation as $\bar f_n^{(1),\chi}= \left(- \partial_{\epsilon} f_n^0 \right) \langle \delta g_n^{(11),\chi} \rangle $ where the chiral chemical potential is defined as
%
%\be
%\left<...\right>_\chi = \frac{ \sum_n \int [dk_z]  \left(-\partial_{\epsilon} f_n^0 \right) (...)}{ \sum_n \int [dk_z]  \left(-\partial_{\epsilon} f_n^0 \right)}
%\ee
%

Using the same approach, we now check for a NL version of the  chiral anomaly by constructing a continuity equation with the 
NL distribution function. For the second-order NL chiral charge pumping equation, we need  the first order NDF. 
The first order NDF can be calculated to be
\be \label{linear_dist}
%f_n^{\chi} \approx 
f_n^{(1),\chi}=\Big( -e \tau_{\omega} E_z v_{z,n}^\chi  +
\kappa \dfrac{\tau_{\omega}}{\tau_v} \langle \delta
g_n^{(1),\chi} \rangle_\chi \Big)\left(-\partial_\epsilon
f_n^{0}\right).
\ee
Here, we have defined %the linear chiral chemical potential
$\langle \delta g_n^{(1),\chi} \rangle_\chi = -e \tau_{v,\omega} E_z \langle v_{z,n}^\chi \rangle_\chi$, with 
$\left<...\right>_\chi = [\sum_n \int [dk_z]  \left(-\partial_{\epsilon} f_n^0 \right) (...)]/ [\sum_n \int [dk_z]  \left(-\partial_{\epsilon} f_n^0 \right)]$ denoting
 the average over all the electronic states at the Fermi level. Additionally, we have used $\tau_\omega=1/(1- i \omega \tau)$, $\tau_{v,\omega}=1/(1- i \omega \tau_v)$ and $\kappa=({\tau_v/\tau} -1)$. Integrating the collisionless Boltzmann transport equation over all the states in the DC limit ($\omega=0$) we obtain,  
\be \label{N_2}
 \frac{\partial {\cal N}^{(2),\chi}}{\partial t} = \frac{e^2 E_z^2 \tau }{4 \pi^2 l_B^2} \bigg[ \left( {\mathcal I}^\chi + {\mathcal C}_{22}^\chi \right)  
 + \kappa \frac{\mathcal{C}_{11}^\chi}{{\mathcal C}_{01}^\chi} {\mathcal C}_{12}^\chi \bigg].
\ee
Here, we have defined
\begin{subequations}
\bea
\mathcal{I}^\chi &&= \sum_n \int dk_z \frac{\partial v_{z,n}^\chi}{\hbar \partial k_z} \left( \partial_{\epsilon} f_n^0\right), \label{I^chi}
\\
\mathcal{C}_{lm}^\chi &&= \sum_n \int dk_z \left(v_{z,n}^\chi \right)^l \left(\partial_{\epsilon}^m f_n^0 \right). \label{C_lm}
\eea
\end{subequations}
In the coefficient $\mathcal{C}_{lm}^\chi$, the first subscript denotes the power of the magnetic band velocity, and the second subscript denotes the order of the derivative of the Fermi function with respect to the energy.
We find that ${\mathcal{I}^\chi + {\mathcal C}_{22}^\chi} = 0$, along with  $\mathcal{C}_{12}^\chi = 0$ and consequently, $\partial {\cal N}^{(2),\chi}/\partial t =0$. This establishes the significant result that there is no NL chiral anomaly that is second order in the electric field strength. Thus, all chiral anomaly related NL transport phenomena in WSM involve the linear chiral anomaly in combination with some other impact of the electric field. 

{\it Nonlinear longitudinal conductivity:---}
Having established that the NL chiral anomaly vanishes, we show that the NL longitudinal conductivity is determined only by the intra-node scattering contributions [see Appendix \ref{app_NDF_calc} for more details]. The corresponding NDF is given by 
%$f_n^{(2),\chi}= \left(- \partial_{\epsilon} f_n^0 \right) \delta g_n^{(21),\chi} +\left(- \partial_{\epsilon}^2 f_n^0 \right) \delta g_n^{(22),\chi}$. 
%We can calculate the local equilibrium parts of the NDF using the particle number conservation in second order of field from the Boltzmann equation. 
%Substituting the ansatz in Eq.~\eqref{BTE}, and following the method developed in Ref.~\cite{deng_PRL2019_posi}, %the non-equilibrium distribution function $\propto |{\bm E}|^2$ reads as follows
%
\be \label{NDF} 
f_n^{(2),\chi}  = \frac{e^2  \tau_{\omega} \tau_{2\omega} E_z^2}{\hbar} \frac{\partial}{\partial{k_z}}  \left[v_{z,n}^\chi \left(\partial_{\epsilon} f_n^0 \right)\right]~,
\ee
where $\tau_{2 \omega}=\tau/(1-i \tau 2 \omega)$. 
Using the NDF in Eq.~(\ref{NDF}), the $2\omega$ component of the longitudinal current
density can be expressed as
\be \label{SHG_cur}
j_z^{2\omega} = -e {\mathfrak D} \sum_{{\chi=\pm 1},n=0}^{n=n_c} \int [dk_z] v_{z,n}^\chi  f_n^{(2),\chi}.
\ee
Here, $n_c$ is the number of filled (empty) LLs in the conduction (valence) band, and it is specified by $n_c= \text{int} \left[\frac{(\mu/\hbar \omega_c)^2}{2 \alpha (1-t_\chi^2)} \right]$. Evaluating Eq.~\eqref{SHG_cur}, yields 
\bea \label{sigma}
\sigma_{zzz}(2\omega) & =& -\frac{e^3 \tau_{\omega} \tau_{2\omega}}{4 \pi^2 \hbar^2 l_B^2 }\sum_{\chi} \mathfrak{I}^\chi,~~{\rm where} \nonumber \\
\mathfrak{I}^\chi &=& - \sum_n \int \big( \partial_{k_z} v_{z,n}^\chi \big) \big( \partial_{k_z} \epsilon_n^\chi \big)
\partial_{\epsilon} f_n^0  dk_z. \label{J^chi}
%- \sum_{n} \int  \frac{\partial v_{z,n}^\chi}{\partial k_z} \frac{\partial \epsilon_n^\chi}{\partial k_z} \frac{\partial f_n^0}{\partial \epsilon_n^\chi} dk_z.
\eea
In deriving Eq.~\eqref{sigma}, we have used 
$  \int \hbar v_{z,n}^\chi \partial_{k_z} \big[v_{z,n}^\chi \partial_{\epsilon} f_n^0  \big] dk_z = -  \int \big( \partial_{k_z} v_{z,n}^\chi \big) \big( \partial_{k_z} \epsilon_n^\chi \big)
\partial_{\epsilon} f_n^0  dk_z $. Using the LL spectrum in Eq.~\eqref{sigma}, we can now calculate the explicit form of the NL longitudinal conductivity, which is completely independent of the inter-node scattering timescale.

Clearly, the finite frequency NL conductivity is complex. The real part of the NL conductivity can be probed in nonlinear transport measurements. The imaginary part provides the information of second harmonic generation~\cite{bhalla_arxiv2021_second} where the NL optical susceptibility is given by $\chi_{zzz}(2\omega)=\sigma_{zzz}(2 \omega)/(i 2 \omega \epsilon_0)$~\cite{gao_PRB2021_cur} with $\epsilon_0$ being the vacuum permittivity. Furthermore, the conductivity can be extracted in two different limits: i) the transport (dominated by scattering) limit where $\omega \tau \ll 1$ and we get the transport conductivity proportional to $ \tau^2$ and independent of frequency, and ii) the optical or clean limit $\omega \tau \gg 1$, where the NL optical conductivity is proportional to $ 1/\omega^2$ and independent of the scattering time.

%{\bf results for three different regimes}

{\it Quantum oscillations in the NL conductivity:---} 
%{\it Ultra-quantum limit:---}  
In presence of a strong magnetic field, the NL conductivity is expected to show quantum oscillation owing to the discrete LLs. To demonstrate this explicitly we  calculate the contributions to the NL conductivity for each LL in the zero-temperature limit where the derivative of the Fermi function is approximated by the Dirac delta function. For the lowest LL ($n=0$) we calculate
\be
\mathfrak{I}^{\chi}=0~,
\ee
which implies that the chiral LL does not contribute to the longitudinal second-order NL response. This is in contrast to the linear response regime, where the chiral LL have a finite contribution in the longitudinal conductivity~\cite{deng_PRL2019_posi,deng_PRB2019_modu,das_PRR2020_chiral}. However, this actually can be seen from Eq.~\eqref{sigma}. Since the velocity $v_{z,0}^\chi$ has no $k_z$ dependence, the integrand of Eq.~\eqref{sigma} itself is zero. The importance of this result can be appreciated from the fact that for a large enough magnetic field where only the lowest LL ($n=0$) is filled, known as the ultra-quantum limit, the NL longitudinal conductivity of the WSM will vanish. More specifically, the ultra-quantum regime is specified by $B> B_{\rm max}\equiv \frac{\mu^2}{2 \hbar e v_F^2 \alpha (1-t_\chi^2)}$, and we predict that the NL longitudinal conductivity vanishes in this regime. This is also consistent with the vanishing NL chiral anomaly shown earlier. 

The contributions from each of the higher LLs to the NL longitudinal conductivity, in the low-temperature limit (for $\mu>0$) is given by
\be \label{SHG_int_tilted} 
\mathfrak{I}^{\chi} = \sum_n
\hbar v_F^2  2|n| (\alpha \hbar \omega_c)^2  \left[  1/{\mathcal E}_{n,\tilde k_{z0}^+ }^3 -  1/{\mathcal E}_{n,\tilde k_{z0}^-}^3 \right] ~~ n \geq 1~.
\ee
Here, the momentum cuts $\tilde k_{z0}^\pm$ on the Fermi surface ($\mu$) are obtained from Eq.~\eqref{momentum_cut} after substituting $k\to k_F=\mu/\hbar v_F$.
We note that Eq.~\eqref{SHG_int_tilted} is independent of chirality and $\mathfrak{I}^{\chi} =0$ for zero tilt ($t_{\parallel,\chi}=0$). The latter can be seen from the fact that for zero tilt velocity, the momentum cuts on the Fermi surface satisfy $\tilde k_{z0}^+=-\tilde k_{z0}^-$ and consequently ${\mathcal E}_{n,\tilde k_{z0}^+ }={\mathcal E}_{n,\tilde k_{z0}^- }$. 
\begin{figure}
    \centering
    \includegraphics[width=\linewidth]{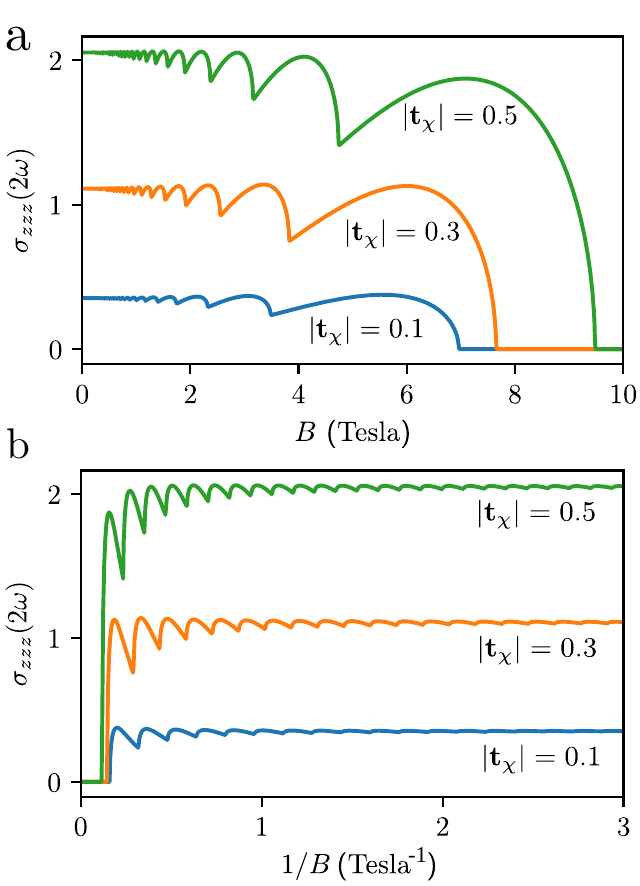}
    \caption{The nonlinear conductivity of a pair of Weyl nodes as a function of a) the magnetic field ($B$) and b) inverse of the magnetic field ($1/B$) for $\mu=20$ meV. The conductivity axis has been scaled by $e^3 \tau_{\omega} \tau_{2\omega} \mu/\hbar^3 \times 10^{-2}$ and we have considered ${\bm t}_\chi=-|{\bm t}_\chi|(\sin \theta, 0, \cos \theta)$ with $\theta=\pi/6$. We note that both the plots complement each other.  Clearly, the NL  conductivity vanishes in the ultra-quantum limit (right [left] side of the panel (a) [(b)]), and it becomes constant in the semiclassical regime. 
}
    \label{fig_3}
\end{figure}

The result presented above is for a single Weyl node, and contributions from different Weyl nodes need to be added to obtain 
the total NL response. This is where the different symmetries of the WSM, play a significant role in determining the total NL response from all Weyl nodes. 
We find that when ${\bm t}_+=-{\bm t}_-$, the total NL conductivity, after summing over nodes of opposite chirality ($\mathfrak{J}^{+} + \mathfrak{J}^{-}$), is identically zero. 
Using the explicit expression of $\mathcal{E}_{n,k_z}$ along with Eq.~\eqref{momentum_cut} in Eq.~\eqref{SHG_int_tilted}, we can simplify that ${\mathfrak I}^{\chi} \propto 1/(A- t_{\parallel,\chi})^{3/2} - 1/(A+ t_{\parallel,\chi})^{3/2}$ where $A$ is a quantity independent of the sign of tilt. It is clear from this simplified form that if we add contributions from opposite tilt, the total contribution becomes zero.
%{\textcolor{red}{Add some more detail here, else the argument is not at all clear to the reader}. 
This is also consistent with the fact that we have ${\bm t}_{+}=-{\bm t}_{-}$ in WSM with SIS~\cite{Gao_opt21, chen_PRB2019_optical}. Therefore, the total NL response is only non-zero when the Weyl nodes of opposite chirality have the same tilt orientation and in that case, the total contribution is double of a single Weyl node. 
In a TRS invariant WSM, a minimum of four nodes are allowed and the nodes with same chirality have opposite tilt orientation. In that case, since Eq.~\eqref{SHG_int_tilted} is chirality independent, the total NL response from the same chirality nodes will be opposite to each other and the total response will be identically zero. Therefore, we conclude that the NL conductivity discussed in this paper is non-zero only in WSM where both the TRS and SIS are broken. 
%We also add here that tilt in the opposite chirality nodes can be positive or negative (${\bm w}_{\chi}$ can have the same or opposite sign to that of $v_F$) and in these two cases, the NL conductivity will have opposite sign.

The oscillating nature of the NL conductivity as a function of the applied magnetic field is shown in Fig.~\ref{fig_3}(a). As expected the NL conductivity increases with the increase in tilt. Depending on the strength of the magnetic field three key features can be inferred from the plot. In the small magnetic field (semiclassical) regime with a large number of filled LLs, the NL conductivity is almost independent of the magnetic field. In the ultra-quantum regime for a large magnetic field, $B>B_{\rm max}$ to be precise, the NL conductivity vanishes. %This is expected from the fact that for a fixed chemical potential $\mu$, $B_{\rm max}$ keeps the Fermi level only in the lowest ($n=0$) LL. 
In the intermediate range of the magnetic field, we see pronounced quantum oscillation feature. The usual periodic nature of the quantum oscillations in 1/$B$ can be clearly seen in Fig.~\ref{fig_3}(b). We calculate the oscillation period to be
\be
\Delta(1/B)= 2 \alpha (1-t_\chi^2) e \hbar (v_F/\mu)^2.
\ee 
At its core, this period of quantum oscillations arises from the corresponding period in the density of states, and it manifests in linear as well as in NL magnetotransport. 
\begin{figure}
    \centering
    \includegraphics[width=\linewidth]{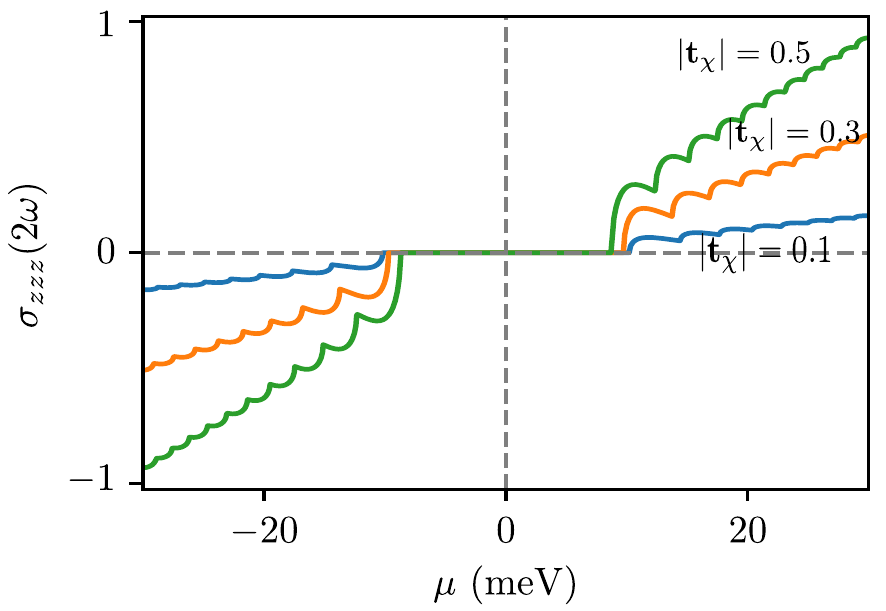}
    \caption{The nonlinear conductivity of a pair of Weyl nodes as a function of chemical potential $\mu$ for $B=2$ Tesla. The NL conductivity has been scaled by $e^3 \tau_{\omega} \tau_{2\omega}/(\hbar^2 \tau)$. We find that the NL conductivity has opposite sign for the chemical potential lying in the valence band and in the conduction band. While the tilt breaks the  particle-hole symmetry, the difference between the magnitude of the NL conductivity in the conduction band and the valance band is not significant. The tilt parameter used here is the same as that in Fig.~\ref{fig_2}, and we have chosen $\tau=10^{-12}$ sec.}
    \label{fig_4}
\end{figure}
We show the chemical potential dependence of the NL conductivity  in Fig.~\ref{fig_4}. For a fixed value of the magnetic field, we find quantum oscillation in the intermediate range of chemical potential and the NL conductivity vanishes for a small value of $\mu$ (ultra-quantum regime). For a large value of $\mu$, when several LLs are filled (semiclassical regime), the NL conductivity has a linear $\mu$ dependence. We emphasize here that the presence of a finite tilt velocity, breaks the particle-hole symmetry and consequently the NL conductivity of the valence band side ($\mu<0$) is different from that in the conduction band side. The NL conductivity of the valence band side has the same form as Eq.~\eqref{SHG_int_tilted}, however, the Fermi surface cuts are modified to
\be 
\tilde k_{z0}^{v, \pm} = \frac{k_F t_{\parallel,\chi} \pm \sqrt{  (1-t_\chi^2) \left( k_F^2 - 2 |n| \alpha^3 /l_B^2 \right)+ k_F^2 t_{\parallel,\chi}^2  }}{\left(1-t_\chi^2\right)},
\ee 
where $k_F=|\mu|/\hbar v_F$. Using this we can see that the NL conductivity has a different sign depending on $\mu$ lying in the conduction band or in the valance band. This can be clearly seen in Fig.~\ref{fig_4}. 

%%%%%{\bf linear in tilt result}

{\it Semiclassical limit of the NL conductivity:---} In the semiclassical limit where many LLs are filled, we can assume the LL index $n$ to be a continuous variable and replace the $\sum_{n}$ by $\int_0^{n_c} dn$ in {Eq.~\eqref{sigma}. Using this integration trick with appropriate limits, it is straightforward to calculate the semiclassical limit of Eq.~\eqref{sigma}. Unfortunately, it has quite a complicated form which obfuscates physical insights. 
However, the limiting case of small tilt velocity is more tractable, and offers useful insights. So we retain the tilt velocity in Eq.~\eqref{SHG_int_tilted} only up to linear order, 
and then take the semiclassical limit. Following this, we approximate $\tilde k_{z0}^{\pm} \approx \pm \sqrt{ \left( k_F^2 - 2 |n| /l_B^2 \right)} - k_F t_{\parallel,\chi}$, and consequently ${\mathcal E}_{n,\tilde k_{z0}^\pm} \approx [\mu^2  \mp 2 \mu  t_{\parallel,\chi}\sqrt{\mu^2  - 2|n| (\hbar \omega_c)^2}]^{1/2} $. Using these simplifications, we find that $\mathfrak{I}^\chi$ in the small tilt velocity approximation is given by
%
%\textcolor{red}{(This needs to be revised and updated)}
\be \label{small_tilt}
\mathfrak{I}^{\chi} \approx \sum_n \hbar v_F^2  2|n| (\hbar \omega_c)^2 \frac{ 6 t_{\parallel, \chi}}{\mu^4} \sqrt{\mu^2 - 2|n| (\hbar \omega_c)^2} ~.
\ee
%
%We note that Eq.~(\ref{small_tilt}) for the NL conductivity 
The approximate expression of the NL conductivity is obtained using Eq.~\eqref{small_tilt} in Eq.~\eqref{sigma}. We find that the NL conductivity, exhibits quantum oscillation behaviour in $1/B$, due to  the LL crossing the chemical potential [see the term $\sqrt{\mu^2 - 2|n| (\hbar \omega_c)^2}$ in Eq.~\eqref{small_tilt}], with periodicity $\Delta(1/B)= 2e \hbar (v_F/\mu)^2$. %which 
This period is identical to that found in the linear magnetoconductance of a
WSM without any tilt velocity~\cite{gao_PRB2021_cur}. Within the linear order tilt approximation, the maximum filled LL index is simplified as $n_c= \text{int}\left[\mu^2/{2 \hbar^2 \omega_c^2} \right]$. 

Using these, it is straightforward to obtain 
$\mathfrak{J}^\chi_{\rm SC} =  \frac{4}{5\hbar} l_B^2  t_{\parallel,\chi} \mu$ in the semiclassical regime. Consequently, the semiclassical NL conductivity is given by, 
\be \label{sigma_semclas}
\sigma_{zzz}^{\rm SC}(2\omega)= - \frac{ e^3 \tau_\omega \tau_{2\omega}}{ \pi^2 \hbar^3} \dfrac{ \mu}{5} \sum_\chi   t_{\parallel,\chi} ~.
\ee 
We find that the NL conductivity in the semiclassical regime is i) $B$-independent, and ii) it varies linearly with $\mu$. 
%These are the key findings of this paper. 
The first observation is quite remarkable, and 
this is also consistent with the more general plot of Fig.~\ref{fig_3}. To understand this better, let's take the extreme limit of zero magnetic fields. In the $B \to 0$ limit, 
the longitudinal NL conductivity should be identical to the NL Drude conductivity, specified by  $\sigma_{zzz}(2\omega) = -e^3 \tau_\omega \tau_{2 \omega}/\hbar \sum_\chi \int [d{\bm k}] v_z \partial_{k_z} (v_z f')$~\cite{sodemann_PRL2015_quan,lahiri_PRB2022_nonlin}. Evaluating this expression, we find that it is identical to the magnetoconductivity obtained in Eq.~\eqref{sigma_semclas}, establishing the consistency of our calculations.  %This shows that the magnetic field has no impact on the longitudinal NL conductivity in the semiclassical regime. 
Since the NL Drude conductivity can only be finite in materials in which both the TRS and SIS is broken, this also helps in understanding the symmetry imposition (absence of both TRS and SIS) for having a finite NL longitudinal magnetoconductivity in tilted WSM.

{\it NL conductivity in multi-Weyl semimetal:---}
Having demonstrated longitudinal NL magnetoconductivity in tilted WSM, we now show their presence in multi-WSM~\cite{dantas_JHEP2018_mag, menon_PRB2021_chiral}. %Although a comprehensive study of the multi-Weyl class is a subject of a separate paper, which we keep for a future direction, we illustrate the central idea by deriving the general expression of the NL conductivity.
The multi-WSM possesses nodes with chirality that have a non-zero integer value. The WSM can be considered to be a special case with the chirality of $\pm 1$. The low energy model Hamiltonian of multi-WSM is given by~\cite{mukherjee_PRB2018_dop,nandy_PRB2021_chiral}
\bea \nn
\mathcal{H}_\chi^\nu & =& \chi [\alpha_\nu (\hbar k_\perp)^\nu \{ \cos (\nu \phi) \sigma_x + \sin (\nu \phi) \sigma_y\} 
+  \hbar v k_z\sigma_z ] \\ \label{ham_multi} 
& & +~ \hbar {\bm w}_\chi \cdot {\bm k}~ \sigma_0.
\eea
Here, $k_\perp=\sqrt{k_x^2 + k_y^2}$ is the perpendicular momentum, $\tan \phi=k_y/k_x$, $\nu$ denotes the chiral charge and $\alpha_\nu$ is a  material-dependent parameter. 
%For example, $\alpha_1$ and $\alpha_2$ are the Fermi velocity and the inverse of the mass respectively for single and double WSM.
The LL problem of a system described by Eq.~\eqref{ham_multi} in the absence of tilt (${\bm w}_\chi = 0$) has been earlier explored in Refs.~\cite{li_PRB2016_weyl, gupta_PLA2019_novel}. Here, we generalize the LL spectrum for the tilted type-I multi-WSM (see Appendix \ref{app_LL_multi} for details). For simplicity, we assume that the tilt is parallel to $\bm B$ {\it i.e.}, ${\bm w}_\chi=(0,0,w_{\parallel,\chi})=v(0,0,t_{\parallel,\chi})$. We find the LL spectrums to be
\be \label{LL_multi}
\epsilon_n^\chi =
\begin{cases}
-\chi \hbar v k_z + \hbar w_{\parallel,\chi} k_z & ~~~\text{for }~~ n < \nu, \\[5pt] 
s \sqrt{\mathcal{F}(n,\alpha_\nu,B)+ \epsilon_z^2} + \hbar w_{\parallel,\chi} k_z &~~\text{ for}~~ n \geq \nu.
\end{cases}
\ee 
Here, we have defined  $\mathcal{F}(n,\alpha_\nu,B)=  n(n-1)...(n-\nu+1) \omega_{\nu}^{2} $, $\omega_\nu = \alpha_\nu (\sqrt{2} \hbar / l_B)^\nu$, $\epsilon_z=\hbar v k_z$, and $s=\pm$ where $+ (-)$ represents energy for conduction (valence) band side. 
The lowest LLs are chiral, disperse linearly and they are $\nu$-fold degenerate.

\begin{figure}
    \centering
    \includegraphics[width=\linewidth]{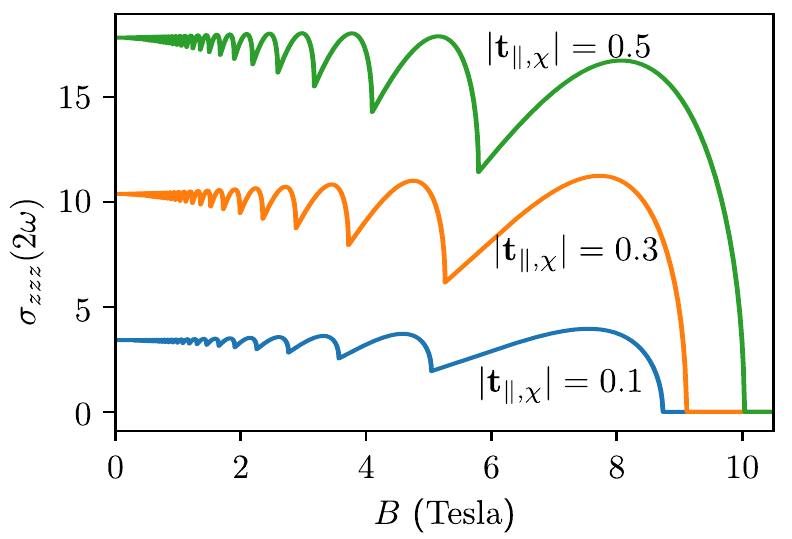}
    \caption{The nonlinear conductivity of a double Weyl node as a function of the magnetic field $\bm B$ for $\mu = 0.15$ meV. The NL conductivity axis has been scaled by $e^3 \tau_{\omega} \tau_{2\omega} \mu/ \hbar^3 $ and we have considered ${\bm t}_\chi=-|{\bm t}_{\parallel,\chi}|(0, 0, 1)$. Here, we have used $\alpha_2 = 0.4$ eV {\AA}/$\hbar^2$, and $v = 0.37 $ eV {\AA}/$\hbar$ \cite{nandy_PRB2021_chiral}.} %\textcolor{blue}{KD:please plot for three tilts and use $t_\parallel$ for multi-Weyl.}}
    \label{fig_5}
\end{figure}

The NL conductivity in the quantum oscillation regime in multi-WSM has the same form as Eq.~\eqref{sigma}, with the modified $\mathfrak{J}^\chi$ specified by
\be 
\mathfrak{J}^\chi = \sum_n \hbar v^2 \mathcal{F}(n,\alpha_\nu,B) \left[1/\mathcal{E}_{n,\tilde k_{z0}^+}^3-1/\mathcal{E}_{n,\tilde k_{z0}^-}^3  \right]~~~\text{$n \geq \nu$}~.
\ee
Here, $\mathcal{E}_{n,\tilde k_{z0}}= \Big[\mathcal{F}(n,\alpha_\nu,B)+(\hbar v {\tilde k}_{z0})^2\Big]^{1/2}$. The momentum corresponding to the Fermi surface are specified by 
\bea \nn
\tilde k_{z0}^{\pm} &=&  \Big[ \pm \sqrt{\left(1-t_{\parallel,\chi}^2 \right) \left[k_F^2- \mathcal{F}(n,\alpha_\nu,B)/(\hbar v)^2 \right]+ k_F^2 t_{\parallel,\chi}^2} 
\\ & & - k_F t_{\parallel,\chi} \Big]\left(1-t_{\parallel,\chi}^2\right)^{-1},
\eea
Here, we have used $k_F=\mu/(\hbar v)$. To calculate the total NL current, we sum over all the occupied non-degenerate ($n \geq \nu$) LLs in Eq.~\eqref{sigma} where the maximum filled LL index $n_c$ is specified by $\mathcal{F}(n,\alpha_\nu,B) = \mu^2/(1-t_{\parallel,\chi}^2)$. 
We have shown the oscillating NL conductivity of double-WSM in Fig.~\ref{fig_5} as a function of the magnetic field with $n_c = {\rm int} \left[ 1/2+ \sqrt{1/4+  \mu^2 /[\omega^2_2 (1-t_{\parallel,\chi}^2)]} \right]$. We find that the multi-WSM shows features in the NL conductivity, which are very similar to those calculated for WSM. Specifically, we find that the NL conductivity vanishes in the ultra-quantum limit, oscillates in the intermediate regime and saturates in the semiclassical regime. Furthermore, the fundamental symmetry constraints and the relative tilt orientation of the multi-Weyl nodes of opposite chirality to get non-zero $\sigma_{zzz}(2\omega)$ in multi-WSM are also the same as those in WSM. However, we note that the double WSM can withstand the oscillation in NL conductivity for a relatively large magnetic field value or very low chemical potential as compared to the WSM. 

%\section{Discussions}
{\it Discussions and conclusion:---} To summarize, we have demonstrated the existence of finite NL longitudinal  magnetoconductivity in type-I WSM with quantized LLs. We show that the NL longitudinal conductivity is finite only in WSM where both the TRS and SIS are broken. We argue that in WSM,  there is no nonlinear chiral anomaly, where the chiral charge pumped is proportional to the square of the electric field strength. We demonstrate that the NL conductivity is solely  determined by intra-node scatterings and i) vanishes in the ultra-quantum limit where only the chiral LL are occupied, ii) displays quantum oscillations in 1/$B$ with a tilt velocity-dependent period, and iii) becomes $B$ independent in the semiclassical regime, reducing to the NL generalization of the Drude conductivity. Further analysis reveals that similar physics is also at play in the broader class of tilted multi-WSM. 
The NL conductivity discussed in this paper will manifest through NL resistance in the case of transport experiments while through second harmonic generation in optical experiments. Due to the decisive dependence of the NL conductivity on the tilt orientation, we believe that our study will play pivotal role in determining tilt configuration of real Weyl materials.

{\it Note:---} During the final preparation stage of this manuscript, we came across Ref.~\cite{zeng_arxiv2022_quantum} by Zeng {\it et. al.} where the nonlinear planar Hall effect 
is explored. 
%due to linear chiral anomaly has been discussed.  

\section*{ACKNOWLEDGEMENTS}
We acknowledge the Science and Engineering Research Board (SERB), and the Department of Science and Technology (DST) of the Government of India for financial support. K.D. and S.D. 
thank IIT Kanpur for the research fellowship. We sincerely thank Pushpendra Yadav for useful discussions.

\appendix
\section{Derivation of second order nonequilibrium distribution function} 
\label{app_NDF_calc}

In this Appendix, we provide the intermediate steps for calculating the second-order NDF. To calculate this, first we need the linear order distribution function. The latter has been calculated earlier in several Refs.~\cite{deng_PRL2019_posi, deng_PRB2019_modu, shao_JPCM2021_long}. With an ansatz of the form $f_n^{(1),\chi} =  \delta g_n^{(1),\chi} \left(- \partial_{\epsilon} f_n^0 \right)$, the Boltzmann equation upto linear order in electric field can be constructed as
\be \label{linearized_BTE}
-i \omega \delta g_n^{(1),\chi} -e E_z {v}_{z,n}^\chi = - \frac{\delta g_n^{(1),\chi} - \delta \Bar{g}_n^{(1),\chi}}{\tau} - \frac{\delta \Bar{g}_n^{(1),\chi}}{\tau_v} .
\ee
Using the particle number conservation within each node, it is straightforward to calculate the linear distribution function which is given in Eq.~\eqref{linear_dist} of the main text. To calculate the second-order NDF, we extend the same formalism to include the quadratic electric field effects. Since the second-order NDF is expected to contain both the first and second derivative of the Fermi function, we consider an ansatz of the form 
\be \label{f2}
f_n^{(2),\chi} =  \delta g_n^{(21),\chi} \left(- \partial_{\epsilon} f_n^0 \right) + \delta g_n^{(22),\chi} \left(- \partial_{\epsilon}^2 f_n^0 \right).
\ee
Here, the first superscript ($i$) in $\delta g^{(ij),\chi}_n$ denotes the electric field dependence and the second superscript ($j$) denotes the order of energy derivative on the Fermi function. To begin with, we consider that the second-order distribution function changes the local equilibrium and the local part has the following form
\be
\bar f_n^{(2),\chi} =  \delta \bar g_n^{(21),\chi} \left(- \partial_{\epsilon} f_n^0 \right) + \delta \bar {\bar g}_n^{(22),\chi} \left(- \partial_{\epsilon}^2 f_n^0 \right). ~~~ \label{bar_f2}
\ee
Here, $\delta \bar g_n^{(21),\chi} \equiv \langle \delta g_n^{(21),\chi}\rangle_\chi$, and the definition of average is same as defined in the main text. However, for the $\delta \bar {\bar g}_n^{(22),\chi} \equiv \langle \! \langle  \delta g_n^{(22),\chi}\rangle \! \rangle_\chi$ we define 
\be
\langle \! \langle ... \rangle \! \rangle_\chi = \frac{ \sum_n \int [dk_z]  \left(-\partial^2_\epsilon f_n^0 \right) (...) }{ \sum_n \int [dk_z]  \left(-\partial^2_\epsilon f_n^0 \right)}.
\ee
Now, using Eqs.~(\ref{f2}, \ref{bar_f2}) in the Boltzmann equation, we obtain the nonlinear version of it as
\begin{widetext}
\bea \label{BTE_2}
& & -2i\omega \left[ \delta g_n^{(21),\chi} \left(- \partial_{\epsilon} f_n^0 \right) + \delta g_n^{(22),\chi} \left(- \partial_{\epsilon}^2 f_n^0 \right) \right] - \frac{e E_z}{\hbar} \partial_{k_z}  \left[ \delta g_n^{(1),\chi} \left(-\partial_\epsilon f_n^0 \right)\right]   =
- \frac{1}{\tau} \left[\delta g_n^{(21),\chi} - \delta \Bar{g}_n^{(21),\chi} \right] \left(-\partial_\epsilon f_n^0 \right) \nn \\
& & -\frac{1}{\tau} \left[\delta g_n^{(22),\chi} -\delta \Bar{\bar g}_n^{(22),\chi }\right] \left(-\partial_{\epsilon}^2 f_n^0 \right)  - \frac{1}{\tau_v}  \left[ \delta \Bar{ g}_n^{(21),\chi}  \left(-\partial_\epsilon f_n^0 \right) + \delta {\Bar {\bar g}}_n^{(22),\chi}  \left(-\partial_{\epsilon}^2 f_n^0 \right) \right]~.
\eea  
Now, we integrate both sides of the above equation with $ \sum_n \int [dk_z]$ and divide by $ \sum_n \int [d k_z] (-\partial_\epsilon f_n^0)$. As the intranode scattering does not alter the number of particles within each node, so all the terms $\propto 1/\tau$ on the right-hand side will get canceled. Consequently, we are left with the equation 
\bea \nn
& & -2i \omega \delta \bar{g}_n^{(21),\chi}  - 2i\omega {\mathcal D}\int_{n,k_z}  \delta g_n^{(22),\chi} \left(- \partial_{\epsilon}^2 f_n^0 \right)  -
e E_z {\mathcal D} \int_{n,k_z}  {v}_{z,n}^\chi \delta g_{n}^{(1),\chi} \left(-\partial_\epsilon^2 f_n^0 \right)  -  e E_z/{\hbar} {\mathcal D}  \int_{n,k_z} \partial_{k_z}  \delta g_n^{(1),\chi}  \left(-\partial_\epsilon f_n^0 \right) \\ \label{int_bte}
& & = 
  - \frac{1}{\tau_v} \delta \Bar{g}_n^{(21),\chi } - \frac{1}{\tau_v} {\mathcal D}  \int_{n,k_z} \delta {\Bar {\bar g}}_n^{(22),\chi}  \left(-\partial_\epsilon^2 f_n^0 \right)~.
\eea 
\end{widetext}
Here, we have defined $ \sum_n \int [dk_z] \equiv \int_{n,k_z} $ and $1/{\mathcal D} \equiv \sum_n \int [d k_z] (-\partial_\epsilon f_n^0)$ for brevity. Now, from our definition of $\delta \bar g_n^{(21),\chi}$, in the above equation we identify
\be \label{bar_g_21}
\delta \bar{g}_n^{(21),\chi} = \frac{e \tau_{v,2\omega} E_z /{\hbar} \sum_n \int [dk_z] \left(-\partial_\epsilon f_n^0 \right)  \partial_{k_z} \delta g_n^{(1),\chi} }{ \sum_n \int [dk_z] \left(-\partial_\epsilon f_n^0 \right)} .
\ee
With this, the integrated Boltzmann equation, Eq.~\eqref{int_bte}, reduces to
\bea \nn
 & & 2i\omega  \int_{n,k_z}  \delta g_n^{(22),\chi}  (-\partial_{\epsilon}^2 f_n^0 ) + e E_z  \int_{n,k_z}  {v}_{z,n}^\chi \delta g_{n}^{(1),\chi} (-\partial_\epsilon^2 f_n^0) 
\\
&&= 
\frac{1}{\tau_v}  \int_{n,k_z} \delta \bar {\bar g}_n^{(22),\chi}  (-\partial_\epsilon^2 f_n^0)~.
\eea
Dividing the above equation by $ \sum_n \int [dk_z] \left(- \partial_\epsilon^2 f_n^0  \right)$ we obtain the other part of the local distribution function as
\be \label{bar_g_22}
\delta \bar {\bar g }_n^{(22),\chi} = \frac{e \tau_{v,2\omega} E_z \sum_n \int [dk_z]  {v}_{z,n}^\chi \delta g_n^{(1),\chi} (-\partial_\epsilon^2 f_n^0)}{ \sum_n \int [dk_z] (-\partial_\epsilon^2 f_n^0)}.
\ee
Finally, we use Eqs.~\eqref{bar_g_21} and \eqref{bar_g_22} in Eq.~\eqref{BTE_2} to obtain the second-order NL distribution function. The two components are calculated to be
%
%\begin{subequations}\label{NDF_final}
%\bea \label{NDF2a}
%\delta g_n^{(21),\chi} \left(-\partial_\epsilon f_n^0 \right) =&&   e \tau_{2\omega} E_z/{\hbar} ~\hat{\bm z} \cdot { \nabla}_{\bm k} (\delta g_n^{(1),\chi}) \left(-\partial_\epsilon f_n^0 \right) + \tau_{2\omega}  \frac{\kappa}{\tau_v} \delta \bar{g}_n^{(21),\chi} \left(-\partial_\epsilon f_n^0 \right), \\
%\delta g_n^{(22),\chi} \left(-\partial_\epsilon^2 f_n^0 \right) =&&   e \tau_{2\omega} E_z v_{z,n}^\chi (\delta g_n^{(1),\chi}) \left(-\partial_\epsilon^2 f_n^0 \right) + \tau_{2\omega}  \frac{\kappa}{\tau_v} \delta {\bar g}_n^{(22),\chi} \left(-\partial_\epsilon^2 f_n^0 \right). \label{NDF2b}
%\eea 
%\end{subequations}
%
\begin{subequations}\label{NDF_final}
\bea \label{NDF2a}
\delta g_n^{(21),\chi} = e \tau_{2\omega} E_z/\hbar{}  \partial_{k_z} \delta g_n^{(1),\chi} + \tau_{2\omega}  \frac{\kappa}{\tau_v} \delta \bar{g}_n^{(21),\chi},~~~~~~ 
\\ \label{NDF2b}
\delta g_n^{(22),\chi}  =   e \tau_{2\omega} E_z v_{z,n}^\chi \delta g_n^{(1),\chi}  + \tau_{2\omega}  \frac{\kappa}{\tau_v} \delta \bar {\bar g}_n^{(22),\chi}. ~~~~~~
\eea 
\end{subequations}
Now using the NDF, we calculate the $2\omega$ component of the longitudinal current density, defined as 
\bea 
j_z(2\omega) =&& -e \mathfrak{D} \sum_{n,\chi} \int [dk_z] v_{z,n}^\chi \left[  \delta g_n^{(21),\chi} \left(- \partial_\epsilon f_n^0 \right) \right. \nn \\
&& \left. + \delta g_n^{(22),\chi} \left(- \partial_\epsilon^2 f_n^0 \right)  \right].
\eea 
After a little algebra, we find that the NL longitudinal conductivity has the following form
\be \label{sigma_full}
\begin{aligned} 
\sigma_{zzz}(2\omega)  =  -\frac{e^3 \tau_{\omega} \tau_{2\omega}}{4 \pi^2 l_B^2 } \sum_\chi \Bigg[ \frac{\mathfrak{I}^\chi}{\hbar^2} +  \frac{\kappa }{\tau_v} \Bigg \{ \frac{\mathcal{C}_{11}^\chi}{\mathcal{C}_{01}^\chi} \big(\tau_{v,\omega}   \mathcal{C}_{22}^\chi    
\\
+\tau_{v,2\omega}  \mathcal{I}^\chi \big) + \tau_{v,2\omega} \frac{\mathcal{C}_{12}^{\chi}}{ \mathcal{C}_{02}^\chi}\left(\mathcal{C}_{22}^\chi + \kappa \frac{ \tau_{v,\omega} }{\tau_v}  \frac{\mathcal{C}_{11}^\chi}{\mathcal{C}_{01}^\chi} \mathcal{C}_{12}^\chi \right)  \Bigg \} \Bigg]. 
\end{aligned}
\ee
This is the exact expression of the NL conductivity discussed in this paper.
Note that, $\sigma_{zzz}(2\omega)$ contains all the NL chiral anomaly coefficients defined in the main text in Eqs.~\eqref{I^chi} and \eqref{C_lm}. Using the results of the main text, ${\mathcal C}_{12}^\chi=0$ and ${\mathcal C}_{22}^\chi=- {\mathcal I}^\chi$ it is evident that the chiral anomaly contribution to the NL magnetoconductivity becomes identically zero in the DC limit ($\omega=0$). Furthermore, in case of AC transport limit, where we can consider $\omega \tau_v \ll 1$, which is the interest of this paper, we can ignore all the contributions from the inter-node scattering since in this limit $\tau_{v,\omega} \approx \tau_{v,2\omega}$.

\section{Calculation of Landau levels in multi-Weyl semimetals} \label{app_LL_multi}

In this Appendix, we present the details of the LL calculation of the tilted multi-WSMs~\cite{gupta_PLA2019_novel}. For that, we write the Hamiltonian given in Eq.~\eqref{ham_multi} as 
\be
{\cal H}^\nu_\chi =\chi [\alpha_\nu \{ (\hbar {\hat k}_{-})^\nu \sigma_{+} + (\hbar {\hat k}_{+})^\nu \sigma_{-} \}   +  \hbar v k_z\sigma_z ] + \hbar {\bm w}_\chi \cdot {\bm k} \sigma_0, 
\ee
where ${\hat k}_{\pm} = {\hat k}_x \pm i {\hat k}_y$, $\sigma_{\pm} = \frac{1}{2} (\sigma_x  \pm i \sigma_y )$. We choose the gauge potential to be ${\bm A}=(-By,0,0)$ for the magnetic field along the $z$-direction, parallel to the tilt velocity ${\bm w}_\chi=(0,0,w_{\parallel,\chi})=v(0,0,t_{\parallel,\chi})$. Consequently, the translation symmetry remains invariant along the $x$- and $z$-direction. Hence, we look for the solution of the form ${\cal H}_\chi^\nu \Psi = \epsilon_n \Psi$, where $\Psi= \psi(y) e^{i\hbar (k_x x + k_z z)}$. With such plane wave basis the Hamiltonian is modified as
%
%\be
%{\cal H}^\nu_\chi = \chi 
% \begin{pmatrix}
% \hbar v {\hat k}_z + \chi \hbar w_{\parallel,\chi} {\hat k}_z  & \alpha_\nu (\hbar {\hat k}_x - eB y - i  \hbar {\hat k}_y)^\nu \\[2ex]
% \alpha_\nu (\hbar {\hat k}_x - eB y + i  \hbar {\hat k}_y)^\nu & - \hbar v  {\hat k}_z + \chi \hbar w_{\parallel,\chi} {\hat k}_z 
%\end{pmatrix}
%\ee
%
\be
{\cal H}^\nu_\chi = \chi 
\begin{pmatrix}
\hbar v k_z + \chi \hbar w_{\parallel,\chi} k_z    & \alpha_\nu (\hbar k_x - eB y - i \hbar {\hat k}_y)^\nu 
\\[2ex]
\alpha_\nu (\hbar k_x - eB y + i \hbar {\hat k}_y)^\nu & - \hbar v k_z + \chi \hbar w_{\parallel,\chi} k_z  
\end{pmatrix}
\ee
%
%In absence of tilt this problem has been solved in Ref.~\cite{gupta_PLA2019_novel}. 
To diagonalize the above Hamiltonian, we introduce a new variable ${\tilde y} = (y/l_B - k_x l_B)$, and subsequently the creation and annihilation operators ${\hat a}^\dagger = 1/\sqrt{2} ({\tilde y} - \partial_{\tilde y})$ and ${\hat a} = 1/\sqrt{2} ({\tilde y} + \partial_{\tilde y})$, satisfying the commutation relation $[{\hat a} , {\hat a}^\dagger]=1$. Using these we obtain 
\bea \label{ham_multi_matrix form}
{\cal H}^\nu_\chi = \chi 
\begin{pmatrix}
\hbar v k_z + \chi \hbar w_{\parallel,\chi} k_z  &  (-1)^\nu \omega_\nu (\hat{a})^\nu 
\\[2ex]
(-1)^\nu \omega_\nu (\hat{a}^{\dagger})^\nu & - \hbar v k_z + \chi \hbar w_{\parallel,\chi} k_z %
\end{pmatrix},~~
\eea 
with $\omega_\nu = \alpha_\nu (\sqrt{2} \hbar/l_B )^\nu$. One can obtain the LLs for the above Hamiltonian using the spinor i) $\psi(y) = [a_\nu \psi_{n - \nu}~~ b_\nu \psi_{\nu}]^T$ when $n \geq \nu$, and ii) $\psi(y) = [0 ~~ \psi_{0}]^T$ when $n<\nu$. Here, $\psi_\nu$ are the usual harmonic oscillator wave functions, $a_\nu$ and $b_\nu$ are the normalisation constants. We have provided the LL spectrums in Eq.~\eqref{LL_multi}.

From Eq.~\eqref{LL_multi}, we notice that similar to the WSM, for the multi-WSM also the tilt introduces a constant velocity parallel to the magnetic field, which is crucial for non-zero NL response in the system. Furthermore, the topological charge of the WSM ($\nu$) manifests through the LL spectrum when subjected to a strong magnetic field, which in turn modifies the NL magnetoconductivity in multi-WSM.

\bibliography{NL_MT_LL_WM}

\begin{thebibliography}{75}%
\makeatletter
\providecommand \@ifxundefined [1]{%
 \@ifx{#1\undefined}
}%
\providecommand \@ifnum [1]{%
 \ifnum #1\expandafter \@firstoftwo
 \else \expandafter \@secondoftwo
 \fi
}%
\providecommand \@ifx [1]{%
 \ifx #1\expandafter \@firstoftwo
 \else \expandafter \@secondoftwo
 \fi
}%
\providecommand \natexlab [1]{#1}%
\providecommand \enquote  [1]{``#1''}%
\providecommand \bibnamefont  [1]{#1}%
\providecommand \bibfnamefont [1]{#1}%
\providecommand \citenamefont [1]{#1}%
\providecommand \href@noop [0]{\@secondoftwo}%
\providecommand \href [0]{\begingroup \@sanitize@url \@href}%
\providecommand \@href[1]{\@@startlink{#1}\@@href}%
\providecommand \@@href[1]{\endgroup#1\@@endlink}%
\providecommand \@sanitize@url [0]{\catcode `\\12\catcode `\$12\catcode
  `\&12\catcode `\#12\catcode `\^12\catcode `\_12\catcode `\%12\relax}%
\providecommand \@@startlink[1]{}%
\providecommand \@@endlink[0]{}%
\providecommand \url  [0]{\begingroup\@sanitize@url \@url }%
\providecommand \@url [1]{\endgroup\@href {#1}{\urlprefix }}%
\providecommand \urlprefix  [0]{URL }%
\providecommand \Eprint [0]{\href }%
\providecommand \doibase [0]{http://dx.doi.org/}%
\providecommand \selectlanguage [0]{\@gobble}%
\providecommand \bibinfo  [0]{\@secondoftwo}%
\providecommand \bibfield  [0]{\@secondoftwo}%
\providecommand \translation [1]{[#1]}%
\providecommand \BibitemOpen [0]{}%
\providecommand \bibitemStop [0]{}%
\providecommand \bibitemNoStop [0]{.\EOS\space}%
\providecommand \EOS [0]{\spacefactor3000\relax}%
\providecommand \BibitemShut  [1]{\csname bibitem#1\endcsname}%
\let\auto@bib@innerbib\@empty
%</preamble>
\bibitem [{\citenamefont {Wan}\ \emph {et~al.}(2011)\citenamefont {Wan},
  \citenamefont {Turner}, \citenamefont {Vishwanath},\ and\ \citenamefont
  {Savrasov}}]{wan_PRB2011_topo}%
  \BibitemOpen
  \bibfield  {author} {\bibinfo {author} {\bibfnamefont {X.}~\bibnamefont
  {Wan}}, \bibinfo {author} {\bibfnamefont {A.~M.}\ \bibnamefont {Turner}},
  \bibinfo {author} {\bibfnamefont {A.}~\bibnamefont {Vishwanath}}, \ and\
  \bibinfo {author} {\bibfnamefont {S.~Y.}\ \bibnamefont {Savrasov}},\ }\href
  {\doibase 10.1103/PhysRevB.83.205101} {\bibfield  {journal} {\bibinfo
  {journal} {Phys. Rev. B}\ }\textbf {\bibinfo {volume} {83}},\ \bibinfo
  {pages} {205101} (\bibinfo {year} {2011})}\BibitemShut {NoStop}%
\bibitem [{\citenamefont {Burkov}\ and\ \citenamefont
  {Balents}(2011)}]{burkov_PRL2011_weyl}%
  \BibitemOpen
  \bibfield  {author} {\bibinfo {author} {\bibfnamefont {A.~A.}\ \bibnamefont
  {Burkov}}\ and\ \bibinfo {author} {\bibfnamefont {L.}~\bibnamefont
  {Balents}},\ }\href {\doibase 10.1103/PhysRevLett.107.127205} {\bibfield
  {journal} {\bibinfo  {journal} {Phys. Rev. Lett.}\ }\textbf {\bibinfo
  {volume} {107}},\ \bibinfo {pages} {127205} (\bibinfo {year}
  {2011})}\BibitemShut {NoStop}%
\bibitem [{\citenamefont {Hasan}\ \emph {et~al.}(2017)\citenamefont {Hasan},
  \citenamefont {Xu}, \citenamefont {Belopolski},\ and\ \citenamefont
  {Huang}}]{hasan_ARCM2017_disco}%
  \BibitemOpen
  \bibfield  {author} {\bibinfo {author} {\bibfnamefont {M.~Z.}\ \bibnamefont
  {Hasan}}, \bibinfo {author} {\bibfnamefont {S.-Y.}\ \bibnamefont {Xu}},
  \bibinfo {author} {\bibfnamefont {I.}~\bibnamefont {Belopolski}}, \ and\
  \bibinfo {author} {\bibfnamefont {S.-M.}\ \bibnamefont {Huang}},\ }\href
  {\doibase 10.1146/annurev-conmatphys-031016-025225} {\bibfield  {journal}
  {\bibinfo  {journal} {Annual Review of Condensed Matter Physics}\ }\textbf
  {\bibinfo {volume} {8}},\ \bibinfo {pages} {289} (\bibinfo {year}
  {2017})}\BibitemShut {NoStop}%
\bibitem [{\citenamefont {Yan}\ and\ \citenamefont
  {Felser}(2017)}]{yan_ARCM2017_topo}%
  \BibitemOpen
  \bibfield  {author} {\bibinfo {author} {\bibfnamefont {B.}~\bibnamefont
  {Yan}}\ and\ \bibinfo {author} {\bibfnamefont {C.}~\bibnamefont {Felser}},\
  }\href {\doibase 10.1146/annurev-conmatphys-031016-025458} {\bibfield
  {journal} {\bibinfo  {journal} {Annual Review of Condensed Matter Physics}\
  }\textbf {\bibinfo {volume} {8}},\ \bibinfo {pages} {337} (\bibinfo {year}
  {2017})}\BibitemShut {NoStop}%
\bibitem [{\citenamefont {Armitage}\ \emph {et~al.}(2018)\citenamefont
  {Armitage}, \citenamefont {Mele},\ and\ \citenamefont
  {Vishwanath}}]{armitage_RMP2018_weyl}%
  \BibitemOpen
  \bibfield  {author} {\bibinfo {author} {\bibfnamefont {N.~P.}\ \bibnamefont
  {Armitage}}, \bibinfo {author} {\bibfnamefont {E.~J.}\ \bibnamefont {Mele}},
  \ and\ \bibinfo {author} {\bibfnamefont {A.}~\bibnamefont {Vishwanath}},\
  }\href {\doibase 10.1103/RevModPhys.90.015001} {\bibfield  {journal}
  {\bibinfo  {journal} {Rev. Mod. Phys.}\ }\textbf {\bibinfo {volume} {90}},\
  \bibinfo {pages} {015001} (\bibinfo {year} {2018})}\BibitemShut {NoStop}%
\bibitem [{\citenamefont {Lv}\ \emph {et~al.}(2021)\citenamefont {Lv},
  \citenamefont {Qian},\ and\ \citenamefont {Ding}}]{lv_RMP2021_expt}%
  \BibitemOpen
  \bibfield  {author} {\bibinfo {author} {\bibfnamefont {B.~Q.}\ \bibnamefont
  {Lv}}, \bibinfo {author} {\bibfnamefont {T.}~\bibnamefont {Qian}}, \ and\
  \bibinfo {author} {\bibfnamefont {H.}~\bibnamefont {Ding}},\ }\href {\doibase
  10.1103/RevModPhys.93.025002} {\bibfield  {journal} {\bibinfo  {journal}
  {Rev. Mod. Phys.}\ }\textbf {\bibinfo {volume} {93}},\ \bibinfo {pages}
  {025002} (\bibinfo {year} {2021})}\BibitemShut {NoStop}%
\bibitem [{\citenamefont {Nielsen}\ and\ \citenamefont
  {Ninomiya}(1983)}]{nielsen_PLB1981_the}%
  \BibitemOpen
  \bibfield  {author} {\bibinfo {author} {\bibfnamefont {H.}~\bibnamefont
  {Nielsen}}\ and\ \bibinfo {author} {\bibfnamefont {M.}~\bibnamefont
  {Ninomiya}},\ }\href {\doibase https://doi.org/10.1016/0370-2693(83)91529-0}
  {\bibfield  {journal} {\bibinfo  {journal} {Physics Letters B}\ }\textbf
  {\bibinfo {volume} {130}},\ \bibinfo {pages} {389} (\bibinfo {year}
  {1983})}\BibitemShut {NoStop}%
\bibitem [{\citenamefont {Zyuzin}\ and\ \citenamefont
  {Burkov}(2012)}]{zyuzin_PRB2012_topo}%
  \BibitemOpen
  \bibfield  {author} {\bibinfo {author} {\bibfnamefont {A.~A.}\ \bibnamefont
  {Zyuzin}}\ and\ \bibinfo {author} {\bibfnamefont {A.~A.}\ \bibnamefont
  {Burkov}},\ }\href {\doibase 10.1103/PhysRevB.86.115133} {\bibfield
  {journal} {\bibinfo  {journal} {Phys. Rev. B}\ }\textbf {\bibinfo {volume}
  {86}},\ \bibinfo {pages} {115133} (\bibinfo {year} {2012})}\BibitemShut
  {NoStop}%
\bibitem [{\citenamefont {Son}\ and\ \citenamefont
  {Spivak}(2013)}]{son_PRB2013_chiral}%
  \BibitemOpen
  \bibfield  {author} {\bibinfo {author} {\bibfnamefont {D.~T.}\ \bibnamefont
  {Son}}\ and\ \bibinfo {author} {\bibfnamefont {B.~Z.}\ \bibnamefont
  {Spivak}},\ }\href {\doibase 10.1103/PhysRevB.88.104412} {\bibfield
  {journal} {\bibinfo  {journal} {Phys. Rev. B}\ }\textbf {\bibinfo {volume}
  {88}},\ \bibinfo {pages} {104412} (\bibinfo {year} {2013})}\BibitemShut
  {NoStop}%
\bibitem [{\citenamefont {Kim}\ \emph {et~al.}(2013)\citenamefont {Kim},
  \citenamefont {Kim}, \citenamefont {Wang}, \citenamefont {Sasaki},
  \citenamefont {Satoh}, \citenamefont {Ohnishi}, \citenamefont {Kitaura},
  \citenamefont {Yang},\ and\ \citenamefont {Li}}]{kim_PRL2013_dirac}%
  \BibitemOpen
  \bibfield  {author} {\bibinfo {author} {\bibfnamefont {H.-J.}\ \bibnamefont
  {Kim}}, \bibinfo {author} {\bibfnamefont {K.-S.}\ \bibnamefont {Kim}},
  \bibinfo {author} {\bibfnamefont {J.-F.}\ \bibnamefont {Wang}}, \bibinfo
  {author} {\bibfnamefont {M.}~\bibnamefont {Sasaki}}, \bibinfo {author}
  {\bibfnamefont {N.}~\bibnamefont {Satoh}}, \bibinfo {author} {\bibfnamefont
  {A.}~\bibnamefont {Ohnishi}}, \bibinfo {author} {\bibfnamefont
  {M.}~\bibnamefont {Kitaura}}, \bibinfo {author} {\bibfnamefont
  {M.}~\bibnamefont {Yang}}, \ and\ \bibinfo {author} {\bibfnamefont
  {L.}~\bibnamefont {Li}},\ }\href {\doibase 10.1103/PhysRevLett.111.246603}
  {\bibfield  {journal} {\bibinfo  {journal} {Phys. Rev. Lett.}\ }\textbf
  {\bibinfo {volume} {111}},\ \bibinfo {pages} {246603} (\bibinfo {year}
  {2013})}\BibitemShut {NoStop}%
\bibitem [{\citenamefont {Deng}\ \emph
  {et~al.}(2019{\natexlab{a}})\citenamefont {Deng}, \citenamefont {Qi},
  \citenamefont {Ma}, \citenamefont {Shen}, \citenamefont {Wang}, \citenamefont
  {Sheng},\ and\ \citenamefont {Xing}}]{deng_PRL2019_posi}%
  \BibitemOpen
  \bibfield  {author} {\bibinfo {author} {\bibfnamefont {M.-X.}\ \bibnamefont
  {Deng}}, \bibinfo {author} {\bibfnamefont {G.~Y.}\ \bibnamefont {Qi}},
  \bibinfo {author} {\bibfnamefont {R.}~\bibnamefont {Ma}}, \bibinfo {author}
  {\bibfnamefont {R.}~\bibnamefont {Shen}}, \bibinfo {author} {\bibfnamefont
  {R.-Q.}\ \bibnamefont {Wang}}, \bibinfo {author} {\bibfnamefont
  {L.}~\bibnamefont {Sheng}}, \ and\ \bibinfo {author} {\bibfnamefont {D.~Y.}\
  \bibnamefont {Xing}},\ }\href {\doibase 10.1103/PhysRevLett.122.036601}
  {\bibfield  {journal} {\bibinfo  {journal} {Phys. Rev. Lett.}\ }\textbf
  {\bibinfo {volume} {122}},\ \bibinfo {pages} {036601} (\bibinfo {year}
  {2019}{\natexlab{a}})}\BibitemShut {NoStop}%
\bibitem [{\citenamefont {Das}\ and\ \citenamefont
  {Agarwal}(2020)}]{das_PRR2020_thermal}%
  \BibitemOpen
  \bibfield  {author} {\bibinfo {author} {\bibfnamefont {K.}~\bibnamefont
  {Das}}\ and\ \bibinfo {author} {\bibfnamefont {A.}~\bibnamefont {Agarwal}},\
  }\href {\doibase 10.1103/PhysRevResearch.2.013088} {\bibfield  {journal}
  {\bibinfo  {journal} {Phys. Rev. Research}\ }\textbf {\bibinfo {volume}
  {2}},\ \bibinfo {pages} {013088} (\bibinfo {year} {2020})}\BibitemShut
  {NoStop}%
\bibitem [{\citenamefont {Das}\ \emph {et~al.}(2020)\citenamefont {Das},
  \citenamefont {Singh},\ and\ \citenamefont {Agarwal}}]{das_PRR2020_chiral}%
  \BibitemOpen
  \bibfield  {author} {\bibinfo {author} {\bibfnamefont {K.}~\bibnamefont
  {Das}}, \bibinfo {author} {\bibfnamefont {S.~K.}\ \bibnamefont {Singh}}, \
  and\ \bibinfo {author} {\bibfnamefont {A.}~\bibnamefont {Agarwal}},\ }\href
  {\doibase 10.1103/PhysRevResearch.2.033511} {\bibfield  {journal} {\bibinfo
  {journal} {Phys. Rev. Research}\ }\textbf {\bibinfo {volume} {2}},\ \bibinfo
  {pages} {033511} (\bibinfo {year} {2020})}\BibitemShut {NoStop}%
\bibitem [{\citenamefont {Chernodub}\ \emph {et~al.}(2021)\citenamefont
  {Chernodub}, \citenamefont {Ferreiros}, \citenamefont {Grushin},
  \citenamefont {Landsteiner},\ and\ \citenamefont
  {Vozmediano}}]{chernodub_arxiv2021_thermal}%
  \BibitemOpen
  \bibfield  {author} {\bibinfo {author} {\bibfnamefont {M.~N.}\ \bibnamefont
  {Chernodub}}, \bibinfo {author} {\bibfnamefont {Y.}~\bibnamefont
  {Ferreiros}}, \bibinfo {author} {\bibfnamefont {A.~G.}\ \bibnamefont
  {Grushin}}, \bibinfo {author} {\bibfnamefont {K.}~\bibnamefont
  {Landsteiner}}, \ and\ \bibinfo {author} {\bibfnamefont {M.~A.~H.}\
  \bibnamefont {Vozmediano}},\ }\href@noop {} {\enquote {\bibinfo {title}
  {Thermal transport, geometry, and anomalies},}\ } (\bibinfo {year} {2021}),\
  \Eprint {http://arxiv.org/abs/2110.05471} {arXiv:2110.05471
  [cond-mat.mes-hall]} \BibitemShut {NoStop}%
\bibitem [{\citenamefont {Das}\ and\ \citenamefont
  {Agarwal}(2021)}]{das_PRB2021_intrinsic}%
  \BibitemOpen
  \bibfield  {author} {\bibinfo {author} {\bibfnamefont {K.}~\bibnamefont
  {Das}}\ and\ \bibinfo {author} {\bibfnamefont {A.}~\bibnamefont {Agarwal}},\
  }\href {\doibase 10.1103/PhysRevB.103.125432} {\bibfield  {journal} {\bibinfo
   {journal} {Phys. Rev. B}\ }\textbf {\bibinfo {volume} {103}},\ \bibinfo
  {pages} {125432} (\bibinfo {year} {2021})}\BibitemShut {NoStop}%
\bibitem [{\citenamefont {Li}\ \emph {et~al.}(2016{\natexlab{a}})\citenamefont
  {Li}, \citenamefont {He}, \citenamefont {Lu}, \citenamefont {Zhang},
  \citenamefont {Liu}, \citenamefont {Ma}, \citenamefont {Fan}, \citenamefont
  {Shen},\ and\ \citenamefont {Wang}}]{li_NC2016_nega}%
  \BibitemOpen
  \bibfield  {author} {\bibinfo {author} {\bibfnamefont {H.}~\bibnamefont
  {Li}}, \bibinfo {author} {\bibfnamefont {H.}~\bibnamefont {He}}, \bibinfo
  {author} {\bibfnamefont {H.-Z.}\ \bibnamefont {Lu}}, \bibinfo {author}
  {\bibfnamefont {H.}~\bibnamefont {Zhang}}, \bibinfo {author} {\bibfnamefont
  {H.}~\bibnamefont {Liu}}, \bibinfo {author} {\bibfnamefont {R.}~\bibnamefont
  {Ma}}, \bibinfo {author} {\bibfnamefont {Z.}~\bibnamefont {Fan}}, \bibinfo
  {author} {\bibfnamefont {S.-Q.}\ \bibnamefont {Shen}}, \ and\ \bibinfo
  {author} {\bibfnamefont {J.}~\bibnamefont {Wang}},\ }\href
  {http://dx.doi.org/10.1038/ncomms10301} {\bibfield  {journal} {\bibinfo
  {journal} {Nature Communications}\ }\textbf {\bibinfo {volume} {7}},\
  \bibinfo {pages} {10301} (\bibinfo {year} {2016}{\natexlab{a}})}\BibitemShut
  {NoStop}%
\bibitem [{\citenamefont {Xiong}\ \emph {et~al.}(2015)\citenamefont {Xiong},
  \citenamefont {Kushwaha}, \citenamefont {Liang}, \citenamefont {Krizan},
  \citenamefont {Hirschberger}, \citenamefont {Wang}, \citenamefont {Cava},\
  and\ \citenamefont {Ong}}]{xiong_S2015_eviden}%
  \BibitemOpen
  \bibfield  {author} {\bibinfo {author} {\bibfnamefont {J.}~\bibnamefont
  {Xiong}}, \bibinfo {author} {\bibfnamefont {S.~K.}\ \bibnamefont {Kushwaha}},
  \bibinfo {author} {\bibfnamefont {T.}~\bibnamefont {Liang}}, \bibinfo
  {author} {\bibfnamefont {J.~W.}\ \bibnamefont {Krizan}}, \bibinfo {author}
  {\bibfnamefont {M.}~\bibnamefont {Hirschberger}}, \bibinfo {author}
  {\bibfnamefont {W.}~\bibnamefont {Wang}}, \bibinfo {author} {\bibfnamefont
  {R.~J.}\ \bibnamefont {Cava}}, \ and\ \bibinfo {author} {\bibfnamefont
  {N.~P.}\ \bibnamefont {Ong}},\ }\href {\doibase 10.1126/science.aac6089}
  {\bibfield  {journal} {\bibinfo  {journal} {Science}\ }\textbf {\bibinfo
  {volume} {350}},\ \bibinfo {pages} {413} (\bibinfo {year}
  {2015})}\BibitemShut {NoStop}%
\bibitem [{\citenamefont {Weng}\ \emph {et~al.}(2015)\citenamefont {Weng},
  \citenamefont {Fang}, \citenamefont {Fang}, \citenamefont {Bernevig},\ and\
  \citenamefont {Dai}}]{weng_PRX_weyl}%
  \BibitemOpen
  \bibfield  {author} {\bibinfo {author} {\bibfnamefont {H.}~\bibnamefont
  {Weng}}, \bibinfo {author} {\bibfnamefont {C.}~\bibnamefont {Fang}}, \bibinfo
  {author} {\bibfnamefont {Z.}~\bibnamefont {Fang}}, \bibinfo {author}
  {\bibfnamefont {B.~A.}\ \bibnamefont {Bernevig}}, \ and\ \bibinfo {author}
  {\bibfnamefont {X.}~\bibnamefont {Dai}},\ }\href {\doibase
  10.1103/PhysRevX.5.011029} {\bibfield  {journal} {\bibinfo  {journal} {Phys.
  Rev. X}\ }\textbf {\bibinfo {volume} {5}},\ \bibinfo {pages} {011029}
  (\bibinfo {year} {2015})}\BibitemShut {NoStop}%
\bibitem [{\citenamefont {Lv}\ \emph {et~al.}(2015)\citenamefont {Lv},
  \citenamefont {Weng}, \citenamefont {Fu}, \citenamefont {Wang}, \citenamefont
  {Miao}, \citenamefont {Ma}, \citenamefont {Richard}, \citenamefont {Huang},
  \citenamefont {Zhao}, \citenamefont {Chen}, \citenamefont {Fang},
  \citenamefont {Dai}, \citenamefont {Qian},\ and\ \citenamefont
  {Ding}}]{lv_PRX2015_expt}%
  \BibitemOpen
  \bibfield  {author} {\bibinfo {author} {\bibfnamefont {B.~Q.}\ \bibnamefont
  {Lv}}, \bibinfo {author} {\bibfnamefont {H.~M.}\ \bibnamefont {Weng}},
  \bibinfo {author} {\bibfnamefont {B.~B.}\ \bibnamefont {Fu}}, \bibinfo
  {author} {\bibfnamefont {X.~P.}\ \bibnamefont {Wang}}, \bibinfo {author}
  {\bibfnamefont {H.}~\bibnamefont {Miao}}, \bibinfo {author} {\bibfnamefont
  {J.}~\bibnamefont {Ma}}, \bibinfo {author} {\bibfnamefont {P.}~\bibnamefont
  {Richard}}, \bibinfo {author} {\bibfnamefont {X.~C.}\ \bibnamefont {Huang}},
  \bibinfo {author} {\bibfnamefont {L.~X.}\ \bibnamefont {Zhao}}, \bibinfo
  {author} {\bibfnamefont {G.~F.}\ \bibnamefont {Chen}}, \bibinfo {author}
  {\bibfnamefont {Z.}~\bibnamefont {Fang}}, \bibinfo {author} {\bibfnamefont
  {X.}~\bibnamefont {Dai}}, \bibinfo {author} {\bibfnamefont {T.}~\bibnamefont
  {Qian}}, \ and\ \bibinfo {author} {\bibfnamefont {H.}~\bibnamefont {Ding}},\
  }\href {\doibase 10.1103/PhysRevX.5.031013} {\bibfield  {journal} {\bibinfo
  {journal} {Phys. Rev. X}\ }\textbf {\bibinfo {volume} {5}},\ \bibinfo {pages}
  {031013} (\bibinfo {year} {2015})}\BibitemShut {NoStop}%
\bibitem [{\citenamefont {Xu}\ \emph {et~al.}(2015)\citenamefont {Xu},
  \citenamefont {Belopolski}, \citenamefont {Alidoust}, \citenamefont
  {Neupane}, \citenamefont {Bian}, \citenamefont {Zhang}, \citenamefont
  {Sankar}, \citenamefont {Chang}, \citenamefont {Yuan}, \citenamefont {Lee},
  \citenamefont {Huang}, \citenamefont {Zheng}, \citenamefont {Ma},
  \citenamefont {Sanchez}, \citenamefont {Wang}, \citenamefont {Bansil},
  \citenamefont {Chou}, \citenamefont {Shibayev}, \citenamefont {Lin},
  \citenamefont {Jia},\ and\ \citenamefont {Hasan}}]{xu_S2015_disco}%
  \BibitemOpen
  \bibfield  {author} {\bibinfo {author} {\bibfnamefont {S.-Y.}\ \bibnamefont
  {Xu}}, \bibinfo {author} {\bibfnamefont {I.}~\bibnamefont {Belopolski}},
  \bibinfo {author} {\bibfnamefont {N.}~\bibnamefont {Alidoust}}, \bibinfo
  {author} {\bibfnamefont {M.}~\bibnamefont {Neupane}}, \bibinfo {author}
  {\bibfnamefont {G.}~\bibnamefont {Bian}}, \bibinfo {author} {\bibfnamefont
  {C.}~\bibnamefont {Zhang}}, \bibinfo {author} {\bibfnamefont
  {R.}~\bibnamefont {Sankar}}, \bibinfo {author} {\bibfnamefont
  {G.}~\bibnamefont {Chang}}, \bibinfo {author} {\bibfnamefont
  {Z.}~\bibnamefont {Yuan}}, \bibinfo {author} {\bibfnamefont {C.-C.}\
  \bibnamefont {Lee}}, \bibinfo {author} {\bibfnamefont {S.-M.}\ \bibnamefont
  {Huang}}, \bibinfo {author} {\bibfnamefont {H.}~\bibnamefont {Zheng}},
  \bibinfo {author} {\bibfnamefont {J.}~\bibnamefont {Ma}}, \bibinfo {author}
  {\bibfnamefont {D.~S.}\ \bibnamefont {Sanchez}}, \bibinfo {author}
  {\bibfnamefont {B.}~\bibnamefont {Wang}}, \bibinfo {author} {\bibfnamefont
  {A.}~\bibnamefont {Bansil}}, \bibinfo {author} {\bibfnamefont
  {F.}~\bibnamefont {Chou}}, \bibinfo {author} {\bibfnamefont {P.~P.}\
  \bibnamefont {Shibayev}}, \bibinfo {author} {\bibfnamefont {H.}~\bibnamefont
  {Lin}}, \bibinfo {author} {\bibfnamefont {S.}~\bibnamefont {Jia}}, \ and\
  \bibinfo {author} {\bibfnamefont {M.~Z.}\ \bibnamefont {Hasan}},\ }\href
  {\doibase 10.1126/science.aaa9297} {\bibfield  {journal} {\bibinfo  {journal}
  {Science}\ }\textbf {\bibinfo {volume} {349}},\ \bibinfo {pages} {613}
  (\bibinfo {year} {2015})}\BibitemShut {NoStop}%
\bibitem [{\citenamefont {Huang}\ \emph {et~al.}(2015)\citenamefont {Huang},
  \citenamefont {Xu}, \citenamefont {Belopolski}, \citenamefont {Lee},
  \citenamefont {Chang}, \citenamefont {Wang}, \citenamefont {Alidoust},
  \citenamefont {Bian}, \citenamefont {Neupane}, \citenamefont {Zhang},
  \citenamefont {Jia}, \citenamefont {Bansil}, \citenamefont {Lin},\ and\
  \citenamefont {Hasan}}]{huang_NC2015_a}%
  \BibitemOpen
  \bibfield  {author} {\bibinfo {author} {\bibfnamefont {S.-M.}\ \bibnamefont
  {Huang}}, \bibinfo {author} {\bibfnamefont {S.-Y.}\ \bibnamefont {Xu}},
  \bibinfo {author} {\bibfnamefont {I.}~\bibnamefont {Belopolski}}, \bibinfo
  {author} {\bibfnamefont {C.-C.}\ \bibnamefont {Lee}}, \bibinfo {author}
  {\bibfnamefont {G.}~\bibnamefont {Chang}}, \bibinfo {author} {\bibfnamefont
  {B.}~\bibnamefont {Wang}}, \bibinfo {author} {\bibfnamefont {N.}~\bibnamefont
  {Alidoust}}, \bibinfo {author} {\bibfnamefont {G.}~\bibnamefont {Bian}},
  \bibinfo {author} {\bibfnamefont {M.}~\bibnamefont {Neupane}}, \bibinfo
  {author} {\bibfnamefont {C.}~\bibnamefont {Zhang}}, \bibinfo {author}
  {\bibfnamefont {S.}~\bibnamefont {Jia}}, \bibinfo {author} {\bibfnamefont
  {A.}~\bibnamefont {Bansil}}, \bibinfo {author} {\bibfnamefont
  {H.}~\bibnamefont {Lin}}, \ and\ \bibinfo {author} {\bibfnamefont {M.~Z.}\
  \bibnamefont {Hasan}},\ }\href {\doibase 10.1038/ncomms8373} {\bibfield
  {journal} {\bibinfo  {journal} {Nature Communications}\ }\textbf {\bibinfo
  {volume} {6}},\ \bibinfo {pages} {7373} (\bibinfo {year} {2015})}\BibitemShut
  {NoStop}%
\bibitem [{\citenamefont {Zyuzin}\ \emph {et~al.}(2012)\citenamefont {Zyuzin},
  \citenamefont {Wu},\ and\ \citenamefont {Burkov}}]{zyuzin_PRB2012_weyl}%
  \BibitemOpen
  \bibfield  {author} {\bibinfo {author} {\bibfnamefont {A.~A.}\ \bibnamefont
  {Zyuzin}}, \bibinfo {author} {\bibfnamefont {S.}~\bibnamefont {Wu}}, \ and\
  \bibinfo {author} {\bibfnamefont {A.~A.}\ \bibnamefont {Burkov}},\ }\href
  {\doibase 10.1103/PhysRevB.85.165110} {\bibfield  {journal} {\bibinfo
  {journal} {Phys. Rev. B}\ }\textbf {\bibinfo {volume} {85}},\ \bibinfo
  {pages} {165110} (\bibinfo {year} {2012})}\BibitemShut {NoStop}%
\bibitem [{\citenamefont {Chang}\ \emph {et~al.}(2018)\citenamefont {Chang},
  \citenamefont {Singh}, \citenamefont {Xu}, \citenamefont {Bian},
  \citenamefont {Huang}, \citenamefont {Hsu}, \citenamefont {Belopolski},
  \citenamefont {Alidoust}, \citenamefont {Sanchez}, \citenamefont {Zheng},
  \citenamefont {Lu}, \citenamefont {Zhang}, \citenamefont {Bian},
  \citenamefont {Chang}, \citenamefont {Jeng}, \citenamefont {Bansil},
  \citenamefont {Hsu}, \citenamefont {Jia}, \citenamefont {Neupert},
  \citenamefont {Lin},\ and\ \citenamefont {Hasan}}]{chang_PRB2018_magne}%
  \BibitemOpen
  \bibfield  {author} {\bibinfo {author} {\bibfnamefont {G.}~\bibnamefont
  {Chang}}, \bibinfo {author} {\bibfnamefont {B.}~\bibnamefont {Singh}},
  \bibinfo {author} {\bibfnamefont {S.-Y.}\ \bibnamefont {Xu}}, \bibinfo
  {author} {\bibfnamefont {G.}~\bibnamefont {Bian}}, \bibinfo {author}
  {\bibfnamefont {S.-M.}\ \bibnamefont {Huang}}, \bibinfo {author}
  {\bibfnamefont {C.-H.}\ \bibnamefont {Hsu}}, \bibinfo {author} {\bibfnamefont
  {I.}~\bibnamefont {Belopolski}}, \bibinfo {author} {\bibfnamefont
  {N.}~\bibnamefont {Alidoust}}, \bibinfo {author} {\bibfnamefont {D.~S.}\
  \bibnamefont {Sanchez}}, \bibinfo {author} {\bibfnamefont {H.}~\bibnamefont
  {Zheng}}, \bibinfo {author} {\bibfnamefont {H.}~\bibnamefont {Lu}}, \bibinfo
  {author} {\bibfnamefont {X.}~\bibnamefont {Zhang}}, \bibinfo {author}
  {\bibfnamefont {Y.}~\bibnamefont {Bian}}, \bibinfo {author} {\bibfnamefont
  {T.-R.}\ \bibnamefont {Chang}}, \bibinfo {author} {\bibfnamefont {H.-T.}\
  \bibnamefont {Jeng}}, \bibinfo {author} {\bibfnamefont {A.}~\bibnamefont
  {Bansil}}, \bibinfo {author} {\bibfnamefont {H.}~\bibnamefont {Hsu}},
  \bibinfo {author} {\bibfnamefont {S.}~\bibnamefont {Jia}}, \bibinfo {author}
  {\bibfnamefont {T.}~\bibnamefont {Neupert}}, \bibinfo {author} {\bibfnamefont
  {H.}~\bibnamefont {Lin}}, \ and\ \bibinfo {author} {\bibfnamefont {M.~Z.}\
  \bibnamefont {Hasan}},\ }\href {\doibase 10.1103/PhysRevB.97.041104}
  {\bibfield  {journal} {\bibinfo  {journal} {Phys. Rev. B}\ }\textbf {\bibinfo
  {volume} {97}},\ \bibinfo {pages} {041104} (\bibinfo {year}
  {2018})}\BibitemShut {NoStop}%
\bibitem [{\citenamefont {Yang}\ \emph {et~al.}(2021)\citenamefont {Yang},
  \citenamefont {Singh}, \citenamefont {Gaudet}, \citenamefont {Lu},
  \citenamefont {Huang}, \citenamefont {Chiu}, \citenamefont {Huang},
  \citenamefont {Wang}, \citenamefont {Bahrami}, \citenamefont {Xu},
  \citenamefont {Franklin}, \citenamefont {Sochnikov}, \citenamefont {Graf},
  \citenamefont {Xu}, \citenamefont {Zhao}, \citenamefont {Hoffman},
  \citenamefont {Lin}, \citenamefont {Torchinsky}, \citenamefont {Broholm},
  \citenamefont {Bansil},\ and\ \citenamefont {Tafti}}]{yang_PRB2021_non}%
  \BibitemOpen
  \bibfield  {author} {\bibinfo {author} {\bibfnamefont {H.-Y.}\ \bibnamefont
  {Yang}}, \bibinfo {author} {\bibfnamefont {B.}~\bibnamefont {Singh}},
  \bibinfo {author} {\bibfnamefont {J.}~\bibnamefont {Gaudet}}, \bibinfo
  {author} {\bibfnamefont {B.}~\bibnamefont {Lu}}, \bibinfo {author}
  {\bibfnamefont {C.-Y.}\ \bibnamefont {Huang}}, \bibinfo {author}
  {\bibfnamefont {W.-C.}\ \bibnamefont {Chiu}}, \bibinfo {author}
  {\bibfnamefont {S.-M.}\ \bibnamefont {Huang}}, \bibinfo {author}
  {\bibfnamefont {B.}~\bibnamefont {Wang}}, \bibinfo {author} {\bibfnamefont
  {F.}~\bibnamefont {Bahrami}}, \bibinfo {author} {\bibfnamefont
  {B.}~\bibnamefont {Xu}}, \bibinfo {author} {\bibfnamefont {J.}~\bibnamefont
  {Franklin}}, \bibinfo {author} {\bibfnamefont {I.}~\bibnamefont {Sochnikov}},
  \bibinfo {author} {\bibfnamefont {D.~E.}\ \bibnamefont {Graf}}, \bibinfo
  {author} {\bibfnamefont {G.}~\bibnamefont {Xu}}, \bibinfo {author}
  {\bibfnamefont {Y.}~\bibnamefont {Zhao}}, \bibinfo {author} {\bibfnamefont
  {C.~M.}\ \bibnamefont {Hoffman}}, \bibinfo {author} {\bibfnamefont
  {H.}~\bibnamefont {Lin}}, \bibinfo {author} {\bibfnamefont {D.~H.}\
  \bibnamefont {Torchinsky}}, \bibinfo {author} {\bibfnamefont {C.~L.}\
  \bibnamefont {Broholm}}, \bibinfo {author} {\bibfnamefont {A.}~\bibnamefont
  {Bansil}}, \ and\ \bibinfo {author} {\bibfnamefont {F.}~\bibnamefont
  {Tafti}},\ }\href {\doibase 10.1103/PhysRevB.103.115143} {\bibfield
  {journal} {\bibinfo  {journal} {Phys. Rev. B}\ }\textbf {\bibinfo {volume}
  {103}},\ \bibinfo {pages} {115143} (\bibinfo {year} {2021})}\BibitemShut
  {NoStop}%
\bibitem [{\citenamefont {Boyd}(2020)}]{Boyd20}%
  \BibitemOpen
  \bibfield  {author} {\bibinfo {author} {\bibfnamefont {R.}~\bibnamefont
  {Boyd}},\ }\href {https://books.google.co.in/books?id=54vZDwAAQBAJ} {\emph
  {\bibinfo {title} {Nonlinear Optics}}}\ (\bibinfo  {publisher} {Elsevier
  Science},\ \bibinfo {year} {2020})\BibitemShut {NoStop}%
\bibitem [{\citenamefont {Wu}\ \emph {et~al.}(2017)\citenamefont {Wu},
  \citenamefont {Patankar}, \citenamefont {Morimoto}, \citenamefont {Nair},
  \citenamefont {Thewalt}, \citenamefont {Little}, \citenamefont {Analytis},
  \citenamefont {Moore},\ and\ \citenamefont {Orenstein}}]{Wu17}%
  \BibitemOpen
  \bibfield  {author} {\bibinfo {author} {\bibfnamefont {L.}~\bibnamefont
  {Wu}}, \bibinfo {author} {\bibfnamefont {S.}~\bibnamefont {Patankar}},
  \bibinfo {author} {\bibfnamefont {T.}~\bibnamefont {Morimoto}}, \bibinfo
  {author} {\bibfnamefont {N.~L.}\ \bibnamefont {Nair}}, \bibinfo {author}
  {\bibfnamefont {E.}~\bibnamefont {Thewalt}}, \bibinfo {author} {\bibfnamefont
  {A.}~\bibnamefont {Little}}, \bibinfo {author} {\bibfnamefont {J.~G.}\
  \bibnamefont {Analytis}}, \bibinfo {author} {\bibfnamefont {J.~E.}\
  \bibnamefont {Moore}}, \ and\ \bibinfo {author} {\bibfnamefont
  {J.}~\bibnamefont {Orenstein}},\ }\href {\doibase 10.1038/nphys3969}
  {\bibfield  {journal} {\bibinfo  {journal} {Nature Physics}\ }\textbf
  {\bibinfo {volume} {13}},\ \bibinfo {pages} {350} (\bibinfo {year}
  {2017})}\BibitemShut {NoStop}%
\bibitem [{\citenamefont {Ma}\ \emph {et~al.}(2017)\citenamefont {Ma},
  \citenamefont {Xu}, \citenamefont {Chan}, \citenamefont {Zhang},
  \citenamefont {Chang}, \citenamefont {Lin}, \citenamefont {Xie},
  \citenamefont {Palacios}, \citenamefont {Lin}, \citenamefont {Jia},
  \citenamefont {Lee}, \citenamefont {Jarillo-Herrero},\ and\ \citenamefont
  {Gedik}}]{ma_NP2017_dir}%
  \BibitemOpen
  \bibfield  {author} {\bibinfo {author} {\bibfnamefont {Q.}~\bibnamefont
  {Ma}}, \bibinfo {author} {\bibfnamefont {S.-Y.}\ \bibnamefont {Xu}}, \bibinfo
  {author} {\bibfnamefont {C.-K.}\ \bibnamefont {Chan}}, \bibinfo {author}
  {\bibfnamefont {C.-L.}\ \bibnamefont {Zhang}}, \bibinfo {author}
  {\bibfnamefont {G.}~\bibnamefont {Chang}}, \bibinfo {author} {\bibfnamefont
  {Y.}~\bibnamefont {Lin}}, \bibinfo {author} {\bibfnamefont {W.}~\bibnamefont
  {Xie}}, \bibinfo {author} {\bibfnamefont {T.}~\bibnamefont {Palacios}},
  \bibinfo {author} {\bibfnamefont {H.}~\bibnamefont {Lin}}, \bibinfo {author}
  {\bibfnamefont {S.}~\bibnamefont {Jia}}, \bibinfo {author} {\bibfnamefont
  {P.~A.}\ \bibnamefont {Lee}}, \bibinfo {author} {\bibfnamefont
  {P.}~\bibnamefont {Jarillo-Herrero}}, \ and\ \bibinfo {author} {\bibfnamefont
  {N.}~\bibnamefont {Gedik}},\ }\href {\doibase 10.1038/nphys4146} {\bibfield
  {journal} {\bibinfo  {journal} {Nature Physics}\ }\textbf {\bibinfo {volume}
  {13}},\ \bibinfo {pages} {842} (\bibinfo {year} {2017})}\BibitemShut
  {NoStop}%
\bibitem [{\citenamefont {Chan}\ \emph {et~al.}(2017)\citenamefont {Chan},
  \citenamefont {Lindner}, \citenamefont {Refael},\ and\ \citenamefont
  {Lee}}]{chan_PRB2017_photo}%
  \BibitemOpen
  \bibfield  {author} {\bibinfo {author} {\bibfnamefont {C.-K.}\ \bibnamefont
  {Chan}}, \bibinfo {author} {\bibfnamefont {N.~H.}\ \bibnamefont {Lindner}},
  \bibinfo {author} {\bibfnamefont {G.}~\bibnamefont {Refael}}, \ and\ \bibinfo
  {author} {\bibfnamefont {P.~A.}\ \bibnamefont {Lee}},\ }\href {\doibase
  10.1103/PhysRevB.95.041104} {\bibfield  {journal} {\bibinfo  {journal} {Phys.
  Rev. B}\ }\textbf {\bibinfo {volume} {95}},\ \bibinfo {pages} {041104}
  (\bibinfo {year} {2017})}\BibitemShut {NoStop}%
\bibitem [{\citenamefont {K\"onig}\ \emph {et~al.}(2017)\citenamefont
  {K\"onig}, \citenamefont {Xie}, \citenamefont {Pesin},\ and\ \citenamefont
  {Levchenko}}]{konig_PRB2017_photo}%
  \BibitemOpen
  \bibfield  {author} {\bibinfo {author} {\bibfnamefont {E.~J.}\ \bibnamefont
  {K\"onig}}, \bibinfo {author} {\bibfnamefont {H.-Y.}\ \bibnamefont {Xie}},
  \bibinfo {author} {\bibfnamefont {D.~A.}\ \bibnamefont {Pesin}}, \ and\
  \bibinfo {author} {\bibfnamefont {A.}~\bibnamefont {Levchenko}},\ }\href
  {\doibase 10.1103/PhysRevB.96.075123} {\bibfield  {journal} {\bibinfo
  {journal} {Phys. Rev. B}\ }\textbf {\bibinfo {volume} {96}},\ \bibinfo
  {pages} {075123} (\bibinfo {year} {2017})}\BibitemShut {NoStop}%
\bibitem [{\citenamefont {Golub}\ \emph {et~al.}(2017)\citenamefont {Golub},
  \citenamefont {Ivchenko},\ and\ \citenamefont
  {Spivak}}]{golub_JETPL2017_photo}%
  \BibitemOpen
  \bibfield  {author} {\bibinfo {author} {\bibfnamefont {L.~E.}\ \bibnamefont
  {Golub}}, \bibinfo {author} {\bibfnamefont {E.~L.}\ \bibnamefont {Ivchenko}},
  \ and\ \bibinfo {author} {\bibfnamefont {B.~Z.}\ \bibnamefont {Spivak}},\
  }\href {\doibase 10.1134/S0021364017120062} {\bibfield  {journal} {\bibinfo
  {journal} {JETP Letters}\ }\textbf {\bibinfo {volume} {105}},\ \bibinfo
  {pages} {782} (\bibinfo {year} {2017})}\BibitemShut {NoStop}%
\bibitem [{\citenamefont {Sadhukhan}\ and\ \citenamefont
  {Nag}(2021{\natexlab{a}})}]{sadhukhan_prb21_role}%
  \BibitemOpen
  \bibfield  {author} {\bibinfo {author} {\bibfnamefont {B.}~\bibnamefont
  {Sadhukhan}}\ and\ \bibinfo {author} {\bibfnamefont {T.}~\bibnamefont
  {Nag}},\ }\href {\doibase 10.1103/PhysRevB.103.144308} {\bibfield  {journal}
  {\bibinfo  {journal} {Phys. Rev. B}\ }\textbf {\bibinfo {volume} {103}},\
  \bibinfo {pages} {144308} (\bibinfo {year} {2021}{\natexlab{a}})}\BibitemShut
  {NoStop}%
\bibitem [{\citenamefont {Sadhukhan}\ and\ \citenamefont
  {Nag}(2021{\natexlab{b}})}]{sadhukhan_prb21_electronic}%
  \BibitemOpen
  \bibfield  {author} {\bibinfo {author} {\bibfnamefont {B.}~\bibnamefont
  {Sadhukhan}}\ and\ \bibinfo {author} {\bibfnamefont {T.}~\bibnamefont
  {Nag}},\ }\href {\doibase 10.1103/PhysRevB.104.245122} {\bibfield  {journal}
  {\bibinfo  {journal} {Phys. Rev. B}\ }\textbf {\bibinfo {volume} {104}},\
  \bibinfo {pages} {245122} (\bibinfo {year} {2021}{\natexlab{b}})}\BibitemShut
  {NoStop}%
\bibitem [{\citenamefont {de~Juan}\ \emph {et~al.}(2017)\citenamefont
  {de~Juan}, \citenamefont {Grushin}, \citenamefont {Morimoto},\ and\
  \citenamefont {Moore}}]{dejuan_NC2017_quan}%
  \BibitemOpen
  \bibfield  {author} {\bibinfo {author} {\bibfnamefont {F.}~\bibnamefont
  {de~Juan}}, \bibinfo {author} {\bibfnamefont {A.~G.}\ \bibnamefont
  {Grushin}}, \bibinfo {author} {\bibfnamefont {T.}~\bibnamefont {Morimoto}}, \
  and\ \bibinfo {author} {\bibfnamefont {J.~E.}\ \bibnamefont {Moore}},\ }\href
  {\doibase 10.1038/ncomms15995} {\bibfield  {journal} {\bibinfo  {journal}
  {Nature Communications}\ }\textbf {\bibinfo {volume} {8}},\ \bibinfo {pages}
  {15995} (\bibinfo {year} {2017})}\BibitemShut {NoStop}%
\bibitem [{\citenamefont {Yang}\ \emph {et~al.}(2018)\citenamefont {Yang},
  \citenamefont {Burch},\ and\ \citenamefont {Ran}}]{yang_arxiv2018_div}%
  \BibitemOpen
  \bibfield  {author} {\bibinfo {author} {\bibfnamefont {X.}~\bibnamefont
  {Yang}}, \bibinfo {author} {\bibfnamefont {K.}~\bibnamefont {Burch}}, \ and\
  \bibinfo {author} {\bibfnamefont {Y.}~\bibnamefont {Ran}},\ }\href@noop {}
  {\enquote {\bibinfo {title} {Divergent bulk photovoltaic effect in weyl
  semimetals},}\ } (\bibinfo {year} {2018}),\ \Eprint
  {http://arxiv.org/abs/1712.09363} {arXiv:1712.09363 [cond-mat.mes-hall]}
  \BibitemShut {NoStop}%
\bibitem [{\citenamefont {Li}\ \emph {et~al.}(2018)\citenamefont {Li},
  \citenamefont {Jin}, \citenamefont {Tohyama}, \citenamefont {Iitaka},
  \citenamefont {Zhang},\ and\ \citenamefont {Su}}]{li_PRB2018_sec}%
  \BibitemOpen
  \bibfield  {author} {\bibinfo {author} {\bibfnamefont {Z.}~\bibnamefont
  {Li}}, \bibinfo {author} {\bibfnamefont {Y.-Q.}\ \bibnamefont {Jin}},
  \bibinfo {author} {\bibfnamefont {T.}~\bibnamefont {Tohyama}}, \bibinfo
  {author} {\bibfnamefont {T.}~\bibnamefont {Iitaka}}, \bibinfo {author}
  {\bibfnamefont {J.-X.}\ \bibnamefont {Zhang}}, \ and\ \bibinfo {author}
  {\bibfnamefont {H.}~\bibnamefont {Su}},\ }\href {\doibase
  10.1103/PhysRevB.97.085201} {\bibfield  {journal} {\bibinfo  {journal} {Phys.
  Rev. B}\ }\textbf {\bibinfo {volume} {97}},\ \bibinfo {pages} {085201}
  (\bibinfo {year} {2018})}\BibitemShut {NoStop}%
\bibitem [{\citenamefont {Gao}\ and\ \citenamefont {Ge}(2021)}]{Gao_opt21}%
  \BibitemOpen
  \bibfield  {author} {\bibinfo {author} {\bibfnamefont {Y.}~\bibnamefont
  {Gao}}\ and\ \bibinfo {author} {\bibfnamefont {B.}~\bibnamefont {Ge}},\
  }\href {\doibase 10.1364/OE.414524} {\bibfield  {journal} {\bibinfo
  {journal} {Opt. Express}\ }\textbf {\bibinfo {volume} {29}},\ \bibinfo
  {pages} {6903} (\bibinfo {year} {2021})}\BibitemShut {NoStop}%
\bibitem [{\citenamefont {de~Juan}\ \emph {et~al.}(2020)\citenamefont
  {de~Juan}, \citenamefont {Zhang}, \citenamefont {Morimoto}, \citenamefont
  {Sun}, \citenamefont {Moore},\ and\ \citenamefont
  {Grushin}}]{dejuan_PRR2020_diff}%
  \BibitemOpen
  \bibfield  {author} {\bibinfo {author} {\bibfnamefont {F.}~\bibnamefont
  {de~Juan}}, \bibinfo {author} {\bibfnamefont {Y.}~\bibnamefont {Zhang}},
  \bibinfo {author} {\bibfnamefont {T.}~\bibnamefont {Morimoto}}, \bibinfo
  {author} {\bibfnamefont {Y.}~\bibnamefont {Sun}}, \bibinfo {author}
  {\bibfnamefont {J.~E.}\ \bibnamefont {Moore}}, \ and\ \bibinfo {author}
  {\bibfnamefont {A.~G.}\ \bibnamefont {Grushin}},\ }\href {\doibase
  10.1103/PhysRevResearch.2.012017} {\bibfield  {journal} {\bibinfo  {journal}
  {Phys. Rev. Research}\ }\textbf {\bibinfo {volume} {2}},\ \bibinfo {pages}
  {012017} (\bibinfo {year} {2020})}\BibitemShut {NoStop}%
\bibitem [{\citenamefont {Sodemann}\ and\ \citenamefont
  {Fu}(2015)}]{sodemann_PRL2015_quan}%
  \BibitemOpen
  \bibfield  {author} {\bibinfo {author} {\bibfnamefont {I.}~\bibnamefont
  {Sodemann}}\ and\ \bibinfo {author} {\bibfnamefont {L.}~\bibnamefont {Fu}},\
  }\href {\doibase 10.1103/PhysRevLett.115.216806} {\bibfield  {journal}
  {\bibinfo  {journal} {Phys. Rev. Lett.}\ }\textbf {\bibinfo {volume} {115}},\
  \bibinfo {pages} {216806} (\bibinfo {year} {2015})}\BibitemShut {NoStop}%
\bibitem [{\citenamefont {Rostami}\ and\ \citenamefont
  {Polini}(2018)}]{rostami_PRB2018_non}%
  \BibitemOpen
  \bibfield  {author} {\bibinfo {author} {\bibfnamefont {H.}~\bibnamefont
  {Rostami}}\ and\ \bibinfo {author} {\bibfnamefont {M.}~\bibnamefont
  {Polini}},\ }\href {\doibase 10.1103/PhysRevB.97.195151} {\bibfield
  {journal} {\bibinfo  {journal} {Phys. Rev. B}\ }\textbf {\bibinfo {volume}
  {97}},\ \bibinfo {pages} {195151} (\bibinfo {year} {2018})}\BibitemShut
  {NoStop}%
\bibitem [{\citenamefont {Gao}\ \emph {et~al.}(2020)\citenamefont {Gao},
  \citenamefont {Zhang},\ and\ \citenamefont {Zhang}}]{gao_PRB2020_sec}%
  \BibitemOpen
  \bibfield  {author} {\bibinfo {author} {\bibfnamefont {Y.}~\bibnamefont
  {Gao}}, \bibinfo {author} {\bibfnamefont {F.}~\bibnamefont {Zhang}}, \ and\
  \bibinfo {author} {\bibfnamefont {W.}~\bibnamefont {Zhang}},\ }\href
  {\doibase 10.1103/PhysRevB.102.245116} {\bibfield  {journal} {\bibinfo
  {journal} {Phys. Rev. B}\ }\textbf {\bibinfo {volume} {102}},\ \bibinfo
  {pages} {245116} (\bibinfo {year} {2020})}\BibitemShut {NoStop}%
\bibitem [{\citenamefont {Morimoto}\ \emph {et~al.}(2016)\citenamefont
  {Morimoto}, \citenamefont {Zhong}, \citenamefont {Orenstein},\ and\
  \citenamefont {Moore}}]{morimoto_PRB2016_semi}%
  \BibitemOpen
  \bibfield  {author} {\bibinfo {author} {\bibfnamefont {T.}~\bibnamefont
  {Morimoto}}, \bibinfo {author} {\bibfnamefont {S.}~\bibnamefont {Zhong}},
  \bibinfo {author} {\bibfnamefont {J.}~\bibnamefont {Orenstein}}, \ and\
  \bibinfo {author} {\bibfnamefont {J.~E.}\ \bibnamefont {Moore}},\ }\href
  {\doibase 10.1103/PhysRevB.94.245121} {\bibfield  {journal} {\bibinfo
  {journal} {Phys. Rev. B}\ }\textbf {\bibinfo {volume} {94}},\ \bibinfo
  {pages} {245121} (\bibinfo {year} {2016})}\BibitemShut {NoStop}%
\bibitem [{\citenamefont {Zyuzin}\ \emph {et~al.}(2018)\citenamefont {Zyuzin},
  \citenamefont {Silaev},\ and\ \citenamefont
  {Zyuzin}}]{zyuzin_PRB2018_nonlinear}%
  \BibitemOpen
  \bibfield  {author} {\bibinfo {author} {\bibfnamefont {A.~A.}\ \bibnamefont
  {Zyuzin}}, \bibinfo {author} {\bibfnamefont {M.}~\bibnamefont {Silaev}}, \
  and\ \bibinfo {author} {\bibfnamefont {V.~A.}\ \bibnamefont {Zyuzin}},\
  }\href {\doibase 10.1103/PhysRevB.98.205149} {\bibfield  {journal} {\bibinfo
  {journal} {Phys. Rev. B}\ }\textbf {\bibinfo {volume} {98}},\ \bibinfo
  {pages} {205149} (\bibinfo {year} {2018})}\BibitemShut {NoStop}%
\bibitem [{\citenamefont {Zyuzin}(2018)}]{zyuzin_PRB2018_chiral}%
  \BibitemOpen
  \bibfield  {author} {\bibinfo {author} {\bibfnamefont {V.~A.}\ \bibnamefont
  {Zyuzin}},\ }\href {\doibase 10.1103/PhysRevB.98.165205} {\bibfield
  {journal} {\bibinfo  {journal} {Phys. Rev. B}\ }\textbf {\bibinfo {volume}
  {98}},\ \bibinfo {pages} {165205} (\bibinfo {year} {2018})}\BibitemShut
  {NoStop}%
\bibitem [{\citenamefont {Li}\ \emph {et~al.}(2021)\citenamefont {Li},
  \citenamefont {Heinonen}, \citenamefont {Burkov},\ and\ \citenamefont
  {Zhang}}]{li_PRB2021_non}%
  \BibitemOpen
  \bibfield  {author} {\bibinfo {author} {\bibfnamefont {R.-H.}\ \bibnamefont
  {Li}}, \bibinfo {author} {\bibfnamefont {O.~G.}\ \bibnamefont {Heinonen}},
  \bibinfo {author} {\bibfnamefont {A.~A.}\ \bibnamefont {Burkov}}, \ and\
  \bibinfo {author} {\bibfnamefont {S.~S.-L.}\ \bibnamefont {Zhang}},\ }\href
  {\doibase 10.1103/PhysRevB.103.045105} {\bibfield  {journal} {\bibinfo
  {journal} {Phys. Rev. B}\ }\textbf {\bibinfo {volume} {103}},\ \bibinfo
  {pages} {045105} (\bibinfo {year} {2021})}\BibitemShut {NoStop}%
\bibitem [{\citenamefont {Zeng}\ \emph {et~al.}(2022)\citenamefont {Zeng},
  \citenamefont {Nandy},\ and\ \citenamefont {Tewari}}]{zeng_prb2022_chiral}%
  \BibitemOpen
  \bibfield  {author} {\bibinfo {author} {\bibfnamefont {C.}~\bibnamefont
  {Zeng}}, \bibinfo {author} {\bibfnamefont {S.}~\bibnamefont {Nandy}}, \ and\
  \bibinfo {author} {\bibfnamefont {S.}~\bibnamefont {Tewari}},\ }\href
  {\doibase 10.1103/PhysRevB.105.125131} {\bibfield  {journal} {\bibinfo
  {journal} {Phys. Rev. B}\ }\textbf {\bibinfo {volume} {105}},\ \bibinfo
  {pages} {125131} (\bibinfo {year} {2022})}\BibitemShut {NoStop}%
\bibitem [{\citenamefont {{Nag}}\ and\ \citenamefont
  {{Kennes}}(2022)}]{tanay_arxiv22_distinct}%
  \BibitemOpen
  \bibfield  {author} {\bibinfo {author} {\bibfnamefont {T.}~\bibnamefont
  {{Nag}}}\ and\ \bibinfo {author} {\bibfnamefont {D.~M.}\ \bibnamefont
  {{Kennes}}},\ }\href@noop {} {\  (\bibinfo {year} {2022})},\ \Eprint
  {http://arxiv.org/abs/2201.11417} {arXiv:2201.11417 [cond-mat.mes-hall]}
  \BibitemShut {NoStop}%
\bibitem [{\citenamefont {Kharzeev}\ \emph {et~al.}(2018)\citenamefont
  {Kharzeev}, \citenamefont {Kikuchi}, \citenamefont {Meyer},\ and\
  \citenamefont {Tanizaki}}]{kharzeev_PRB2018_giant}%
  \BibitemOpen
  \bibfield  {author} {\bibinfo {author} {\bibfnamefont {D.~E.}\ \bibnamefont
  {Kharzeev}}, \bibinfo {author} {\bibfnamefont {Y.}~\bibnamefont {Kikuchi}},
  \bibinfo {author} {\bibfnamefont {R.}~\bibnamefont {Meyer}}, \ and\ \bibinfo
  {author} {\bibfnamefont {Y.}~\bibnamefont {Tanizaki}},\ }\href {\doibase
  10.1103/PhysRevB.98.014305} {\bibfield  {journal} {\bibinfo  {journal} {Phys.
  Rev. B}\ }\textbf {\bibinfo {volume} {98}},\ \bibinfo {pages} {014305}
  (\bibinfo {year} {2018})}\BibitemShut {NoStop}%
\bibitem [{\citenamefont {Golub}\ and\ \citenamefont
  {Ivchenko}(2018)}]{golub_PRB2018_circ}%
  \BibitemOpen
  \bibfield  {author} {\bibinfo {author} {\bibfnamefont {L.~E.}\ \bibnamefont
  {Golub}}\ and\ \bibinfo {author} {\bibfnamefont {E.~L.}\ \bibnamefont
  {Ivchenko}},\ }\href {\doibase 10.1103/PhysRevB.98.075305} {\bibfield
  {journal} {\bibinfo  {journal} {Phys. Rev. B}\ }\textbf {\bibinfo {volume}
  {98}},\ \bibinfo {pages} {075305} (\bibinfo {year} {2018})}\BibitemShut
  {NoStop}%
\bibitem [{\citenamefont {Gao}\ and\ \citenamefont
  {Zhang}(2021)}]{gao_PRB2021_cur}%
  \BibitemOpen
  \bibfield  {author} {\bibinfo {author} {\bibfnamefont {Y.}~\bibnamefont
  {Gao}}\ and\ \bibinfo {author} {\bibfnamefont {F.}~\bibnamefont {Zhang}},\
  }\href {\doibase 10.1103/PhysRevB.103.L041301} {\bibfield  {journal}
  {\bibinfo  {journal} {Phys. Rev. B}\ }\textbf {\bibinfo {volume} {103}},\
  \bibinfo {pages} {L041301} (\bibinfo {year} {2021})}\BibitemShut {NoStop}%
\bibitem [{\citenamefont {Li}\ \emph {et~al.}(2016{\natexlab{b}})\citenamefont
  {Li}, \citenamefont {Roy},\ and\ \citenamefont
  {Das~Sarma}}]{li_PRB2016_weyl}%
  \BibitemOpen
  \bibfield  {author} {\bibinfo {author} {\bibfnamefont {X.}~\bibnamefont
  {Li}}, \bibinfo {author} {\bibfnamefont {B.}~\bibnamefont {Roy}}, \ and\
  \bibinfo {author} {\bibfnamefont {S.}~\bibnamefont {Das~Sarma}},\ }\href
  {\doibase 10.1103/PhysRevB.94.195144} {\bibfield  {journal} {\bibinfo
  {journal} {Phys. Rev. B}\ }\textbf {\bibinfo {volume} {94}},\ \bibinfo
  {pages} {195144} (\bibinfo {year} {2016}{\natexlab{b}})}\BibitemShut
  {NoStop}%
\bibitem [{\citenamefont {Gupta}(2019)}]{gupta_PLA2019_novel}%
  \BibitemOpen
  \bibfield  {author} {\bibinfo {author} {\bibfnamefont {A.}~\bibnamefont
  {Gupta}},\ }\href {\doibase https://doi.org/10.1016/j.physleta.2019.04.044}
  {\bibfield  {journal} {\bibinfo  {journal} {Physics Letters A}\ }\textbf
  {\bibinfo {volume} {383}},\ \bibinfo {pages} {2339} (\bibinfo {year}
  {2019})}\BibitemShut {NoStop}%
\bibitem [{\citenamefont {Dantas}\ \emph {et~al.}(2018)\citenamefont {Dantas},
  \citenamefont {Pe{\~{n}}a-Benitez}, \citenamefont {Roy},\ and\ \citenamefont
  {Sur{\'o}wka}}]{dantas_JHEP2018_mag}%
  \BibitemOpen
  \bibfield  {author} {\bibinfo {author} {\bibfnamefont {R.~M.~A.}\
  \bibnamefont {Dantas}}, \bibinfo {author} {\bibfnamefont {F.}~\bibnamefont
  {Pe{\~{n}}a-Benitez}}, \bibinfo {author} {\bibfnamefont {B.}~\bibnamefont
  {Roy}}, \ and\ \bibinfo {author} {\bibfnamefont {P.}~\bibnamefont
  {Sur{\'o}wka}},\ }\href {\doibase 10.1007/JHEP12(2018)069} {\bibfield
  {journal} {\bibinfo  {journal} {Journal of High Energy Physics}\ }\textbf
  {\bibinfo {volume} {2018}},\ \bibinfo {pages} {69} (\bibinfo {year}
  {2018})}\BibitemShut {NoStop}%
\bibitem [{\citenamefont {Menon}\ \emph {et~al.}(2021)\citenamefont {Menon},
  \citenamefont {Chattopadhay},\ and\ \citenamefont
  {Basu}}]{menon_PRB2021_chiral}%
  \BibitemOpen
  \bibfield  {author} {\bibinfo {author} {\bibfnamefont {A.}~\bibnamefont
  {Menon}}, \bibinfo {author} {\bibfnamefont {S.}~\bibnamefont {Chattopadhay}},
  \ and\ \bibinfo {author} {\bibfnamefont {B.}~\bibnamefont {Basu}},\ }\href
  {\doibase 10.1103/PhysRevB.104.075129} {\bibfield  {journal} {\bibinfo
  {journal} {Phys. Rev. B}\ }\textbf {\bibinfo {volume} {104}},\ \bibinfo
  {pages} {075129} (\bibinfo {year} {2021})}\BibitemShut {NoStop}%
\bibitem [{\citenamefont {Mukherjee}\ and\ \citenamefont
  {Carbotte}(2018)}]{mukherjee_PRB2018_dop}%
  \BibitemOpen
  \bibfield  {author} {\bibinfo {author} {\bibfnamefont {S.~P.}\ \bibnamefont
  {Mukherjee}}\ and\ \bibinfo {author} {\bibfnamefont {J.~P.}\ \bibnamefont
  {Carbotte}},\ }\href {\doibase 10.1103/PhysRevB.97.045150} {\bibfield
  {journal} {\bibinfo  {journal} {Phys. Rev. B}\ }\textbf {\bibinfo {volume}
  {97}},\ \bibinfo {pages} {045150} (\bibinfo {year} {2018})}\BibitemShut
  {NoStop}%
\bibitem [{\citenamefont {Nandy}\ \emph {et~al.}(2021)\citenamefont {Nandy},
  \citenamefont {Zeng},\ and\ \citenamefont {Tewari}}]{nandy_PRB2021_chiral}%
  \BibitemOpen
  \bibfield  {author} {\bibinfo {author} {\bibfnamefont {S.}~\bibnamefont
  {Nandy}}, \bibinfo {author} {\bibfnamefont {C.}~\bibnamefont {Zeng}}, \ and\
  \bibinfo {author} {\bibfnamefont {S.}~\bibnamefont {Tewari}},\ }\href
  {\doibase 10.1103/PhysRevB.104.205124} {\bibfield  {journal} {\bibinfo
  {journal} {Phys. Rev. B}\ }\textbf {\bibinfo {volume} {104}},\ \bibinfo
  {pages} {205124} (\bibinfo {year} {2021})}\BibitemShut {NoStop}%
\bibitem [{\citenamefont {Nag}\ and\ \citenamefont
  {Nandy}(2020)}]{nag_JPCM2020_magneto}%
  \BibitemOpen
  \bibfield  {author} {\bibinfo {author} {\bibfnamefont {T.}~\bibnamefont
  {Nag}}\ and\ \bibinfo {author} {\bibfnamefont {S.}~\bibnamefont {Nandy}},\
  }\href {\doibase 10.1088/1361-648x/abc310} {\bibfield  {journal} {\bibinfo
  {journal} {Journal of Physics: Condensed Matter}\ }\textbf {\bibinfo {volume}
  {33}},\ \bibinfo {pages} {075504} (\bibinfo {year} {2020})}\BibitemShut
  {NoStop}%
\bibitem [{\citenamefont {Fu}\ and\ \citenamefont
  {Wang}(2022)}]{fu_PRB2022_thermo}%
  \BibitemOpen
  \bibfield  {author} {\bibinfo {author} {\bibfnamefont {L.~X.}\ \bibnamefont
  {Fu}}\ and\ \bibinfo {author} {\bibfnamefont {C.~M.}\ \bibnamefont {Wang}},\
  }\href {\doibase 10.1103/PhysRevB.105.035201} {\bibfield  {journal} {\bibinfo
   {journal} {Phys. Rev. B}\ }\textbf {\bibinfo {volume} {105}},\ \bibinfo
  {pages} {035201} (\bibinfo {year} {2022})}\BibitemShut {NoStop}%
\bibitem [{\citenamefont {{Xiong}}\ \emph {et~al.}(2022)\citenamefont
  {{Xiong}}, \citenamefont {{Honerkamp}}, \citenamefont {{Kennes}},\ and\
  \citenamefont {{Nag}}}]{xiong_arxiv22_understanding}%
  \BibitemOpen
  \bibfield  {author} {\bibinfo {author} {\bibfnamefont {F.}~\bibnamefont
  {{Xiong}}}, \bibinfo {author} {\bibfnamefont {C.}~\bibnamefont
  {{Honerkamp}}}, \bibinfo {author} {\bibfnamefont {D.~M.}\ \bibnamefont
  {{Kennes}}}, \ and\ \bibinfo {author} {\bibfnamefont {T.}~\bibnamefont
  {{Nag}}},\ }\href@noop {} {\  (\bibinfo {year} {2022})},\ \Eprint
  {http://arxiv.org/abs/2202.08610} {arXiv:2202.08610 [cond-mat.mes-hall]}
  \BibitemShut {NoStop}%
\bibitem [{\citenamefont {{Menon}}\ and\ \citenamefont
  {{Basu}}(2019)}]{anirudha_arxiv19_anomalous}%
  \BibitemOpen
  \bibfield  {author} {\bibinfo {author} {\bibfnamefont {A.}~\bibnamefont
  {{Menon}}}\ and\ \bibinfo {author} {\bibfnamefont {B.}~\bibnamefont
  {{Basu}}},\ }\href@noop {} {\bibfield  {journal} {\bibinfo  {journal} {arXiv
  e-prints}\ ,\ \bibinfo {eid} {arXiv:1901.06716}} (\bibinfo {year} {2019})},\
  \Eprint {http://arxiv.org/abs/1901.06716} {arXiv:1901.06716
  [cond-mat.mes-hall]} \BibitemShut {NoStop}%
\bibitem [{\citenamefont {Nag}\ \emph {et~al.}(2020)\citenamefont {Nag},
  \citenamefont {Menon},\ and\ \citenamefont
  {Basu}}]{tanay_prb20_thermoelectric}%
  \BibitemOpen
  \bibfield  {author} {\bibinfo {author} {\bibfnamefont {T.}~\bibnamefont
  {Nag}}, \bibinfo {author} {\bibfnamefont {A.}~\bibnamefont {Menon}}, \ and\
  \bibinfo {author} {\bibfnamefont {B.}~\bibnamefont {Basu}},\ }\href {\doibase
  10.1103/PhysRevB.102.014307} {\bibfield  {journal} {\bibinfo  {journal}
  {Phys. Rev. B}\ }\textbf {\bibinfo {volume} {102}},\ \bibinfo {pages}
  {014307} (\bibinfo {year} {2020})}\BibitemShut {NoStop}%
\bibitem [{\citenamefont {Sonowal}\ \emph {et~al.}(2019)\citenamefont
  {Sonowal}, \citenamefont {Singh},\ and\ \citenamefont
  {Agarwal}}]{sonowal_PRB2019_giant}%
  \BibitemOpen
  \bibfield  {author} {\bibinfo {author} {\bibfnamefont {K.}~\bibnamefont
  {Sonowal}}, \bibinfo {author} {\bibfnamefont {A.}~\bibnamefont {Singh}}, \
  and\ \bibinfo {author} {\bibfnamefont {A.}~\bibnamefont {Agarwal}},\ }\href
  {\doibase 10.1103/PhysRevB.100.085436} {\bibfield  {journal} {\bibinfo
  {journal} {Phys. Rev. B}\ }\textbf {\bibinfo {volume} {100}},\ \bibinfo
  {pages} {085436} (\bibinfo {year} {2019})}\BibitemShut {NoStop}%
\bibitem [{\citenamefont {Das}\ and\ \citenamefont
  {Agarwal}(2019{\natexlab{a}})}]{das_PRB2019_linear}%
  \BibitemOpen
  \bibfield  {author} {\bibinfo {author} {\bibfnamefont {K.}~\bibnamefont
  {Das}}\ and\ \bibinfo {author} {\bibfnamefont {A.}~\bibnamefont {Agarwal}},\
  }\href {\doibase 10.1103/PhysRevB.99.085405} {\bibfield  {journal} {\bibinfo
  {journal} {Phys. Rev. B}\ }\textbf {\bibinfo {volume} {99}},\ \bibinfo
  {pages} {085405} (\bibinfo {year} {2019}{\natexlab{a}})}\BibitemShut
  {NoStop}%
\bibitem [{\citenamefont {Das}\ and\ \citenamefont
  {Agarwal}(2019{\natexlab{b}})}]{das_PRB2019_berry}%
  \BibitemOpen
  \bibfield  {author} {\bibinfo {author} {\bibfnamefont {K.}~\bibnamefont
  {Das}}\ and\ \bibinfo {author} {\bibfnamefont {A.}~\bibnamefont {Agarwal}},\
  }\href {\doibase 10.1103/PhysRevB.100.085406} {\bibfield  {journal} {\bibinfo
   {journal} {Phys. Rev. B}\ }\textbf {\bibinfo {volume} {100}},\ \bibinfo
  {pages} {085406} (\bibinfo {year} {2019}{\natexlab{b}})}\BibitemShut
  {NoStop}%
\bibitem [{\citenamefont {Yu}\ \emph {et~al.}(2016)\citenamefont {Yu},
  \citenamefont {Yao},\ and\ \citenamefont {Yang}}]{yu_PRL2016_pred}%
  \BibitemOpen
  \bibfield  {author} {\bibinfo {author} {\bibfnamefont {Z.-M.}\ \bibnamefont
  {Yu}}, \bibinfo {author} {\bibfnamefont {Y.}~\bibnamefont {Yao}}, \ and\
  \bibinfo {author} {\bibfnamefont {S.~A.}\ \bibnamefont {Yang}},\ }\href
  {\doibase 10.1103/PhysRevLett.117.077202} {\bibfield  {journal} {\bibinfo
  {journal} {Phys. Rev. Lett.}\ }\textbf {\bibinfo {volume} {117}},\ \bibinfo
  {pages} {077202} (\bibinfo {year} {2016})}\BibitemShut {NoStop}%
\bibitem [{\citenamefont {Chen}\ \emph {et~al.}(2019)\citenamefont {Chen},
  \citenamefont {Kutayiah}, \citenamefont {Oladyshkin}, \citenamefont
  {Tokman},\ and\ \citenamefont {Belyanin}}]{chen_PRB2019_optical}%
  \BibitemOpen
  \bibfield  {author} {\bibinfo {author} {\bibfnamefont {Q.}~\bibnamefont
  {Chen}}, \bibinfo {author} {\bibfnamefont {A.~R.}\ \bibnamefont {Kutayiah}},
  \bibinfo {author} {\bibfnamefont {I.}~\bibnamefont {Oladyshkin}}, \bibinfo
  {author} {\bibfnamefont {M.}~\bibnamefont {Tokman}}, \ and\ \bibinfo {author}
  {\bibfnamefont {A.}~\bibnamefont {Belyanin}},\ }\href {\doibase
  10.1103/PhysRevB.99.075137} {\bibfield  {journal} {\bibinfo  {journal} {Phys.
  Rev. B}\ }\textbf {\bibinfo {volume} {99}},\ \bibinfo {pages} {075137}
  (\bibinfo {year} {2019})}\BibitemShut {NoStop}%
\bibitem [{\citenamefont {Shao}\ and\ \citenamefont
  {Yan}(2021)}]{shao_JPCM2021_long}%
  \BibitemOpen
  \bibfield  {author} {\bibinfo {author} {\bibfnamefont {J.}~\bibnamefont
  {Shao}}\ and\ \bibinfo {author} {\bibfnamefont {L.}~\bibnamefont {Yan}},\
  }\href {\doibase 10.1088/1361-648x/abee3e} {\bibfield  {journal} {\bibinfo
  {journal} {Journal of Physics: Condensed Matter}\ }\textbf {\bibinfo {volume}
  {33}},\ \bibinfo {pages} {185704} (\bibinfo {year} {2021})}\BibitemShut
  {NoStop}%
\bibitem [{\citenamefont {Goerbig}\ \emph {et~al.}(2009)\citenamefont
  {Goerbig}, \citenamefont {Fuchs}, \citenamefont {Montambaux},\ and\
  \citenamefont {Pi{\'{e}}chon}}]{goerbig_EPL2009_elec}%
  \BibitemOpen
  \bibfield  {author} {\bibinfo {author} {\bibfnamefont {M.~O.}\ \bibnamefont
  {Goerbig}}, \bibinfo {author} {\bibfnamefont {J.-N.}\ \bibnamefont {Fuchs}},
  \bibinfo {author} {\bibfnamefont {G.}~\bibnamefont {Montambaux}}, \ and\
  \bibinfo {author} {\bibfnamefont {F.}~\bibnamefont {Pi{\'{e}}chon}},\ }\href
  {\doibase 10.1209/0295-5075/85/57005} {\bibfield  {journal} {\bibinfo
  {journal} {{EPL} (Europhysics Letters)}\ }\textbf {\bibinfo {volume} {85}},\
  \bibinfo {pages} {57005} (\bibinfo {year} {2009})}\BibitemShut {NoStop}%
\bibitem [{\citenamefont {S\'ari}\ \emph {et~al.}(2015)\citenamefont {S\'ari},
  \citenamefont {Goerbig},\ and\ \citenamefont {T\ifmmode~\mbox{\H{o}}\else
  \H{o}\fi{}ke}}]{sari_PRB2015_mag}%
  \BibitemOpen
  \bibfield  {author} {\bibinfo {author} {\bibfnamefont {J.}~\bibnamefont
  {S\'ari}}, \bibinfo {author} {\bibfnamefont {M.~O.}\ \bibnamefont {Goerbig}},
  \ and\ \bibinfo {author} {\bibfnamefont {C.}~\bibnamefont
  {T\ifmmode~\mbox{\H{o}}\else \H{o}\fi{}ke}},\ }\href {\doibase
  10.1103/PhysRevB.92.035306} {\bibfield  {journal} {\bibinfo  {journal} {Phys.
  Rev. B}\ }\textbf {\bibinfo {volume} {92}},\ \bibinfo {pages} {035306}
  (\bibinfo {year} {2015})}\BibitemShut {NoStop}%
\bibitem [{\citenamefont {Lukose}\ \emph {et~al.}(2007)\citenamefont {Lukose},
  \citenamefont {Shankar},\ and\ \citenamefont {Baskaran}}]{Lukose07}%
  \BibitemOpen
  \bibfield  {author} {\bibinfo {author} {\bibfnamefont {V.}~\bibnamefont
  {Lukose}}, \bibinfo {author} {\bibfnamefont {R.}~\bibnamefont {Shankar}}, \
  and\ \bibinfo {author} {\bibfnamefont {G.}~\bibnamefont {Baskaran}},\ }\href
  {\doibase 10.1103/PhysRevLett.98.116802} {\bibfield  {journal} {\bibinfo
  {journal} {Phys. Rev. Lett.}\ }\textbf {\bibinfo {volume} {98}},\ \bibinfo
  {pages} {116802} (\bibinfo {year} {2007})}\BibitemShut {NoStop}%
\bibitem [{\citenamefont {Vadnais}\ and\ \citenamefont
  {C\^ot\'e}(2021)}]{vadnais_PRB2021_quan}%
  \BibitemOpen
  \bibfield  {author} {\bibinfo {author} {\bibfnamefont {S.}~\bibnamefont
  {Vadnais}}\ and\ \bibinfo {author} {\bibfnamefont {R.}~\bibnamefont
  {C\^ot\'e}},\ }\href {\doibase 10.1103/PhysRevB.104.144409} {\bibfield
  {journal} {\bibinfo  {journal} {Phys. Rev. B}\ }\textbf {\bibinfo {volume}
  {104}},\ \bibinfo {pages} {144409} (\bibinfo {year} {2021})}\BibitemShut
  {NoStop}%
\bibitem [{\citenamefont {Deng}\ \emph
  {et~al.}(2019{\natexlab{b}})\citenamefont {Deng}, \citenamefont {Duan},
  \citenamefont {Luo}, \citenamefont {Deng}, \citenamefont {Wang},\ and\
  \citenamefont {Sheng}}]{deng_PRB2019_modu}%
  \BibitemOpen
  \bibfield  {author} {\bibinfo {author} {\bibfnamefont {M.-X.}\ \bibnamefont
  {Deng}}, \bibinfo {author} {\bibfnamefont {H.-J.}\ \bibnamefont {Duan}},
  \bibinfo {author} {\bibfnamefont {W.}~\bibnamefont {Luo}}, \bibinfo {author}
  {\bibfnamefont {W.~Y.}\ \bibnamefont {Deng}}, \bibinfo {author}
  {\bibfnamefont {R.-Q.}\ \bibnamefont {Wang}}, \ and\ \bibinfo {author}
  {\bibfnamefont {L.}~\bibnamefont {Sheng}},\ }\href {\doibase
  10.1103/PhysRevB.99.165146} {\bibfield  {journal} {\bibinfo  {journal} {Phys.
  Rev. B}\ }\textbf {\bibinfo {volume} {99}},\ \bibinfo {pages} {165146}
  (\bibinfo {year} {2019}{\natexlab{b}})}\BibitemShut {NoStop}%
\bibitem [{\citenamefont {Zyuzin}(2017)}]{zyuzin17_chiral}%
  \BibitemOpen
  \bibfield  {author} {\bibinfo {author} {\bibfnamefont {V.~A.}\ \bibnamefont
  {Zyuzin}},\ }\href {\doibase 10.1103/PhysRevB.95.245128} {\bibfield
  {journal} {\bibinfo  {journal} {Phys. Rev. B}\ }\textbf {\bibinfo {volume}
  {95}},\ \bibinfo {pages} {245128} (\bibinfo {year} {2017})}\BibitemShut
  {NoStop}%
\bibitem [{\citenamefont {{Bhalla}}\ \emph {et~al.}(2021)\citenamefont
  {{Bhalla}}, \citenamefont {{Das}}, \citenamefont {{Culcer}},\ and\
  \citenamefont {{Agarwal}}}]{bhalla_arxiv2021_second}%
  \BibitemOpen
  \bibfield  {author} {\bibinfo {author} {\bibfnamefont {P.}~\bibnamefont
  {{Bhalla}}}, \bibinfo {author} {\bibfnamefont {K.}~\bibnamefont {{Das}}},
  \bibinfo {author} {\bibfnamefont {D.}~\bibnamefont {{Culcer}}}, \ and\
  \bibinfo {author} {\bibfnamefont {A.}~\bibnamefont {{Agarwal}}},\ }\href@noop
  {} {\  (\bibinfo {year} {2021})},\ \Eprint {http://arxiv.org/abs/2108.04082}
  {arXiv:2108.04082 [cond-mat.mes-hall]} \BibitemShut {NoStop}%
\bibitem [{\citenamefont {Lahiri}\ \emph {et~al.}(2022)\citenamefont {Lahiri},
  \citenamefont {Bhore}, \citenamefont {Das},\ and\ \citenamefont
  {Agarwal}}]{lahiri_PRB2022_nonlin}%
  \BibitemOpen
  \bibfield  {author} {\bibinfo {author} {\bibfnamefont {S.}~\bibnamefont
  {Lahiri}}, \bibinfo {author} {\bibfnamefont {T.}~\bibnamefont {Bhore}},
  \bibinfo {author} {\bibfnamefont {K.}~\bibnamefont {Das}}, \ and\ \bibinfo
  {author} {\bibfnamefont {A.}~\bibnamefont {Agarwal}},\ }\href {\doibase
  10.1103/PhysRevB.105.045421} {\bibfield  {journal} {\bibinfo  {journal}
  {Phys. Rev. B}\ }\textbf {\bibinfo {volume} {105}},\ \bibinfo {pages}
  {045421} (\bibinfo {year} {2022})}\BibitemShut {NoStop}%
\bibitem [{\citenamefont {{Zeng}}\ \emph {et~al.}(2022)\citenamefont {{Zeng}},
  \citenamefont {{Nandy}}, \citenamefont {{Liu}}, \citenamefont {{Tewari}},\
  and\ \citenamefont {{Yao}}}]{zeng_arxiv2022_quantum}%
  \BibitemOpen
  \bibfield  {author} {\bibinfo {author} {\bibfnamefont {C.}~\bibnamefont
  {{Zeng}}}, \bibinfo {author} {\bibfnamefont {S.}~\bibnamefont {{Nandy}}},
  \bibinfo {author} {\bibfnamefont {P.}~\bibnamefont {{Liu}}}, \bibinfo
  {author} {\bibfnamefont {S.}~\bibnamefont {{Tewari}}}, \ and\ \bibinfo
  {author} {\bibfnamefont {Y.}~\bibnamefont {{Yao}}},\ }\href@noop {} {\
  (\bibinfo {year} {2022})},\ \Eprint {http://arxiv.org/abs/2203.01196}
  {arXiv:2203.01196 [cond-mat.mes-hall]} \BibitemShut {NoStop}%
\end{thebibliography}%
\end{document}